\documentclass[letterpaper,twocolumn,10pt]{article}
\usepackage{usenix}
\microtypecontext{spacing=nonfrench}
\usepackage{amsmath}
\usepackage{amssymb}
\usepackage{adjustbox}
\usepackage{array}
\usepackage{longtable}
\usepackage{multirow}
\usepackage{pifont}
\usepackage{tabularx}
\usepackage{tikz}
\usetikzlibrary{arrows.meta,calc,fit,positioning}
\usepackage{booktabs}
\usepackage{threeparttable}
\usepackage{graphicx}
\usepackage{tabularx}
\usepackage{svg}

\usepackage{pgfplots}
\pgfplotsset{compat=1.18}

\providecommand{\Description}[2][]{}

\makeatletter
\patchcmd{\LT@output}
  {\unvbox\z@\copy\LT@foot\vss}
  {\unvbox\z@\copy\LT@foot\vfil}{}{}
\patchcmd{\LT@output}
  {\unvbox\@cclv\copy\LT@foot\vss}
  {\unvbox\@cclv\copy\LT@foot\vfil}{}{}
\makeatother

\newcolumntype{Y}{>{\raggedright\arraybackslash}X}

\begin{document}

\date{}

\title{\Large \bf SoK: When FL Clients Cannot Train -- Delegated Federated Learning for IoT}

\author{
{\rm Hamed Mohammed Alkharsh}$^{1,2}$\thanks{\raggedright Corresponding authors:
\mbox{\{h.m.h.alghamdi2, Aydin.Abadi\}@newcastle.ac.uk}.},
{\rm Aydin Abadi}$^{1}$\footnotemark[1],\\
{\rm Jakub Szelag}$^{1}$,
{\rm Suleiman Muhammad Sabo}$^{1}$,
{\rm Ishak Semuel Beno}$^{1}$\\[8pt]
$^{1}$Cryptography and AI Security Lab,
School of Computing,
Newcastle University, United Kingdom\\[4pt]
$^{2}$Al-Baha University, Saudi Arabia
}

\begin{NoHyper}
\maketitle
\end{NoHyper}

\begin{abstract}
Federated learning (FL) assumes that a client can perform the computation required to produce its own contribution. This assumption is often unrealistic for resource-constrained Internet of Things (IoT) devices, motivating designs that delegate some or all client-side training to more capable parties. We systematize this setting as delegated federated learning (Delegated FL), where data ownership and execution of owner-attributed learning are separated. We develop a framework that examines three dimensions over the same delegated transition: delegate-view privacy, delegated attribution, and constrained-owner feasibility. Our audit of 29 systems reveals a recurring composition gap. Privacy mechanisms protect selected artifacts without necessarily characterizing the accumulated view. Attribution mechanisms establish individual execution relations without necessarily binding the complete delegated transition. Furthermore, computational offloading does not establish whether the remaining workload fits the owner's resource budget. Two controlled studies on Raspberry Pi devices illustrate these gaps. We show that encrypted inputs can coexist with information leakage through an exposed residual and that locally retained labels remain inferable from split-learning gradients. We also show that message authentication can reject tampering without establishing that the executor performed the prescribed computation. Our findings demonstrate why individually established guarantees do not necessarily support the same end-to-end claim and motivate evaluating privacy, attribution, and residual owner cost over the same delegated transition.
\end{abstract}

\section{Introduction}
\label{sec:introduction}

Federated learning (FL) enables multiple data owners to train a shared model without centralizing their raw data~\cite{mcmahan2017communication,konevcny2016federated,kairouz2021advances}. This makes FL well suited to Internet of Things (IoT) settings, where data originate across sensors, wearables, mobile devices, vehicles, and other connected endpoints rather than in a single repository. In such settings, privacy, ownership, or operational requirements may make centralized collection undesirable~\cite{nguyen2021federated,imteaj2021survey}. FL offers a way to learn across these distributed sources while keeping the underlying records with their owners.

FL, however, still assumes that participating clients can perform local training. A client must receive the current model and execute the prescribed learning procedure. It must also maintain the required state and return a contribution over repeated rounds. For resource-constrained IoT endpoints, this workload may exceed the available device resources or deployment budget~\cite{imteaj2021survey,kaur2023federated,nguyen2021federated}. This creates a natural role for delegation. A sensor may rely on a gateway, a wearable on a smartphone, and a drone or vehicle on a nearby edge or MEC node. Mobile devices may likewise rely on more capable edge or cloud infrastructure. In each case, some or all of the learning needed to produce the owner's FL contribution is performed elsewhere, while the resulting contribution remains attributed to that owner~\cite{lim2020federated,han2024orchestrating}. Delegation can therefore extend FL participation to devices for which repeated local training would otherwise be impractical.

This shift in execution creates a new owner--executor boundary. Computation that would normally remain inside the owner's device is now performed by another party, which may gain access to information that would otherwise remain local~\cite{pasquini2021unleashing,rigaki2023survey}. Depending on the design, the executor may receive raw data or observe intermediate representations produced during learning. These observations may become more revealing when linked across rounds or combined through collusion. The executor also controls computation whose result may later be accepted as the owner's federated contribution. Authenticating that result establishes its origin and integrity, but does not by itself establish that the executor performed the authorized learning procedure on the prescribed data and model state~\cite{zhang2020enabling,chen2020training}. Furthermore, offloading training does not guarantee that the owner becomes lightweight. The communication and protection required for delegated execution may introduce additional work at the constrained device. The resulting question is not only where learning runs, but also what the executor can learn and whether its computation can be attributed to the prescribed task. These guarantees must also be considered in relation to the workload that remains with the owner.

We use \emph{delegated federated learning} (Delegated FL) for settings in which a non-owner performs a semantically necessary learning operation whose result determines a federated contribution attributed to the data owner. This includes full training offload as well as partial or split execution in which the owner retains part of the learning procedure and completes the contribution using an executor-produced intermediate~\cite{vepakomma2018split,poirot2019split,gupta2018distributed}. The distinction requires separating several roles that conventional FL often treats as one. The \emph{logical client} is the entity to which the contribution is attributed, while the \emph{data owner} is authorized to contribute the underlying data. The \emph{executor} performs computation used to derive that contribution. An \emph{administrative principal} may control one or more of these roles. They need not coincide. A logical client may, for example, be a service, gateway, or organizational process rather than the physical sensing endpoint~\cite{wang2021field}. Delegated FL is therefore defined by the separation of data authority from execution authority for an owner-attributed contribution.

This separation does not imply that every delegated design hides raw data from the executor. Some systems expose owner records directly to the party performing the delegated computation and therefore do not preserve endpoint data locality. We include such systems only when the wider protocol remains federated across distinct owners or logical clients and the resulting contribution remains attributable to a particular owner rather than to a globally pooled dataset. What remains federated in these cases is the contribution across owners, not necessarily the processing of each owner's data. Furthermore, resource constraints motivate our study but do not define the execution boundary. Systems motivated by other requirements may exhibit the same separation of ownership and execution. Physical placement is also distinct from the trust boundary. A gateway operated by the owner and an independently administered edge or cloud service may perform the same learning operation under different privacy and accountability assumptions.

Against this background, we ask under what conditions a contribution produced fully or partly by a non-owner can be treated as private, owner-attributable, and feasible for the constrained owner. We examine this question along three dimensions. \emph{Delegate-view privacy} concerns what an executor or coalition can infer from the information available across the delegated interaction and over repeated rounds. \emph{Delegated attribution} concerns whether the accepted contribution can be bound to the owner's authorization and prescribed learning procedure. It also examines how the evidence of execution connects the responsible executor to the admission decision. \emph{Constrained-owner feasibility} concerns whether the complete learning and assurance workload that remains with the owner fits its device and deployment budget. These dimensions are distinct, but they interact because moving or protecting computation changes both what the executor can observe and what the owner must still do.

\noindent\textbf{Positioning.} Existing research addresses important parts of this problem, but usually under different boundaries. SplitFed and related work examine partitioned execution and its privacy boundaries~\cite{hukkeri2025comprehensive,thapa2020advancements}, and broader surveys catalogue privacy and robustness attacks and defenses across FL participants~\cite{lyu2024privacy}. Edge-, fog-, and resource-aware FL research focuses primarily on placement and offloading under heterogeneous client capabilities. General verifiable-computation research studies externally performed work~\cite{gennaro2010non}, while confidential-computing and outsourced-ML research examines how computation or data can be protected on external platforms~\cite{mo2024machine,sharma2021confidential,ma2021survey}. Verifiable FL examines integrity properties of federated training and aggregation~\cite{guo2023advancing,xiao2024risefl,rizzini2026sok}. Each literature addresses part of the delegated setting. What remains less clear is whether the guarantees established by these mechanisms apply to the same owner-attributed contribution under the same execution and trust assumptions.

The closest assurance-oriented work is the SoK on verifiable FL by Rizzini et al.~\cite{rizzini2026sok}. Their work systematizes how evidence can establish integrity claims about FL training and aggregation. It distinguishes properties such as data binding, training correctness, and freshness, and permits the data owner and trainer to be distinct roles. However, this separation is one possible role configuration within their verifiability framework, whereas we make it the object of analysis. Our corpus is therefore defined by delegated execution rather than by the presence of a verifiability mechanism. A verifiable-FL system may fall outside our corpus when the owner still trains locally, while a delegated system remains central to ours even when it provides no verifiable-computation mechanism.

The individual security principles underlying our three dimensions are well established. Our contribution lies in examining how these principles apply to the same delegated transition. For example, verification of an outsourced gradient does not necessarily establish that the gradient corresponds to the current training round. Furthermore, protecting that computation does not establish whether the owner can afford the operations required to realize the protection. As we show in Section~\ref{sec:findings}, systems may establish meaningful guarantees without jointly establishing privacy, attribution, and feasibility for the same contribution. The delegated transition provides a common basis for identifying where these guarantees align and where assurance or deployment evidence remains incomplete.

\noindent\textbf{Contributions.} First, we reframe the FL client abstraction for resource-constrained IoT by separating ownership of a federated contribution from execution of the learning that produces it. This perspective shifts the unit of analysis from a physical device that both owns data and trains locally to the owner-attributed contribution and its execution path. On this basis, we define Delegated FL and develop a framework that analyzes delegate-view privacy, delegated attribution, and constrained-owner feasibility over the same delegated transition. Second, we apply this framework to 29 systems through a property and evidence audit that distinguishes what each design supports from how those claims are substantiated. The synthesis identifies five recurring patterns that explain why guarantees established over individual parts of delegated execution do not necessarily establish the corresponding end-to-end claim. Third, we complement the corpus analysis with two controlled studies that illustrate these patterns at the owner--executor boundary. The studies examine information leakage and the limits of message authentication alongside the resource demands that remain with the owner.

\section{Background and Problem Setting}
\label{sec:background}

This section introduces the client-side FL workflow and the execution boundary created when some or all of that workflow is delegated.

\subsection{Federated Client-Side Training}

Federated learning (FL) coordinates model training across distributed data owners without requiring their raw data to be centralized~\cite{mcmahan2017communication,kairouz2021advances,li2020federated}. In a conventional round, a selected client receives the current model, performs a prescribed local learning procedure, and returns a gradient or model update~\cite{bonawitz2019towards}. Keeping raw data on the client does not by itself guarantee privacy, because model updates, intermediate representations, metadata, and repeated protocol observations may still reveal information about the underlying data~\cite{zhu2019deep,pasquini2021unleashing,rigaki2023survey}.

For resource-constrained IoT devices, local training may exceed available computation, memory, energy, communication, or timing budgets~\cite{imteaj2021survey,nguyen2021federated,pfeiffer2023federated}. Client-side computation can instead be transferred in whole or in part to a gateway, edge node, peer device, accelerator, or cloud service~\cite{he2023computation,dong2020offloading}. These arrangements separate the device that owns or contributes the data from the party that performs some of the learning computation.

\subsection{Delegated Client-Side Execution}

\begin{figure*}[!t]
    \centering
    \includegraphics[width=0.7\textwidth]{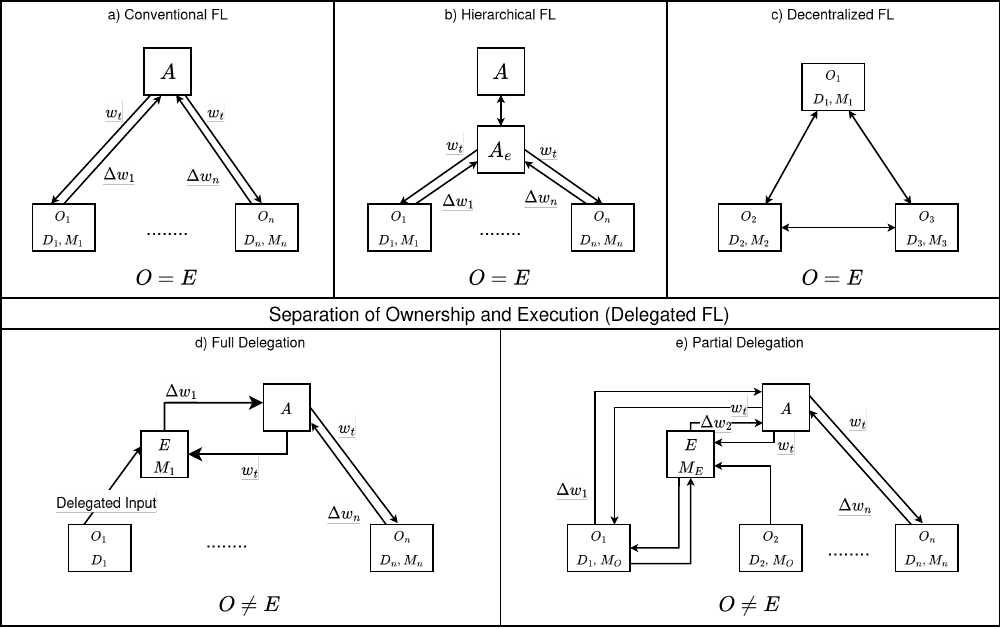}
    \caption{Comparison of federated learning architectures.}
    \label{fig:fl-architectures}

    \vspace{0.5em}
    \begin{minipage}{0.9\textwidth}
        \footnotesize
        \textbf{Notation:}
        \(O_i\), data owner;
        \(E\), executor;
        \(A\), aggregator;
        \(A_e\), edge aggregator;
        \(D_i\), local data;
        \(M\), learning computation;
        \(w_t\), global model;\\
        \(\Delta w_i\), owner-attributed update;
        \(z_i\), intermediate representation;
        \(g_i\), backward signal.
    \end{minipage}
\end{figure*}

We use \emph{Delegated FL} when an executor distinct from the data owner performs a semantically necessary learning operation whose result contributes to an update attributed to that owner. Inclusion follows the execution path that produces the contribution rather than labels such as edge-assisted, split, outsourced, or cloud-supported learning. We exclude operations limited to communication relay, scheduling, resource allocation, aggregation of already-computed updates, inference, or generic infrastructure. Figure~\ref{fig:fl-architectures} contrasts this execution-centered
distinction with conventional, hierarchical, and decentralized FL.
In the latter, the coordination topology may change while the data
owner still performs its local learning computation; under delegation,
ownership and execution of the owner-attributed contribution are separated.

Under \emph{full delegation}, non-owner executors perform all learning operations required to derive the owner-attributed contribution~\cite{dong2020offloading,grafberger2021fedless,zhang2025zkfhed}. The owner may still retain authorization, input preparation or protection, communication, key management, state maintenance, or evidence verification. Under \emph{partial delegation}, the owner retains at least one necessary learning operation and combines it with an executor-produced intermediate result. Split, U-shaped, vertical, peer-assisted, and hierarchical arrangements are distinguished because they expose different artifacts and leave different responsibilities with the owner~\cite{vepakomma2018split,thapa2022splitfed,zhao2023usfl}. Hybrid systems are coded according to the execution mode being evaluated rather than assigned a single unconditional type.

A non-owner executor may still be operated or controlled by the same administrative principal as the owner. For example, an owner-controlled gateway and an independently operated cloud service may perform the same computation while creating different privacy and accountability boundaries. We therefore distinguish logical role separation from administrative control, data visibility, hardware control, and permitted collusion.

\subsection{The Owner--Executor Boundary}

The owner--executor boundary is examined along three dimensions. \emph{Delegate-view privacy} concerns what an executor or coalition can infer from the information available during delegated training, including raw data, labels, intermediate representations, model state, identifiers, timing, and observations linked across rounds. We distinguish what is visible from what is actually inferred or formally allowed to leak, since visibility alone does not establish a successful inference attack~\cite{pasquini2021unleashing,liu2024similarity,kariyappa2023exploit,zhu2025sdar}.

\emph{Delegated attribution} concerns the relations connecting an admitted contribution to the owner's authorization, authorized input, current task and model state, permitted learning procedure, responsible executor, returned result, and admission decision. Authentication establishes message origin and integrity, but it does not establish that the authenticated executor performed the prescribed computation~\cite{krawczyk1997hmac,zhang2022verifiable,rizzini2026sok}. We therefore evaluate work assurance relative to the authorized policy, which may permit sampling, approximation, compression, or variable local steps.

\emph{Constrained-owner feasibility} concerns whether the workload that remains with the owner fits the stated device and deployment budget. Depending on the design, this workload may include preprocessing, cryptography, communication, verification, persistent state, setup, and recovery even when training is delegated. We treat feasibility as a property of the complete residual owner workflow and assess it through computation, peak memory, communication, setup, state, verification, end-to-end latency, and energy where measurable.

These dimensions provide the basis for the delegated-transition model and corpus audit developed in the following sections.
\section{Scope and Corpus Selection}
\label{sec:method}

\noindent\textbf{Scope.} We include systems in which a non-owner performs a semantically necessary learning operation that contributes to an update attributed to the data owner. This covers full delegation and partial delegation, including split execution, but excludes traffic relay, participant scheduling, resource allocation, aggregation of already-computed updates, inference, and generic infrastructure support. Inclusion is determined by the contribution-producing execution path rather than labels such as ``edge-assisted,'' ``split,'' or ``outsourced.'' The motivation for delegation is independent of the execution boundary. We therefore include systems that satisfy this definition even when their original motivation is not constrained IoT deployment. Logical clients also fall within this definition. In OF-RAN~\cite{jijin2022smart}, for example, a resource-constrained client sends its data and model to a seed node, and service nodes train local models and submit updates for aggregation. We include it because the resulting contribution remains attributed to the initiating client, which retains data ownership.

\noindent\textbf{Corpus.} We searched IEEE Xplore between September and October 2025 for work published from 2020 onward. Our queries combined ``federated learning'' with terms for delegation, outsourcing, offloading, proxies, split or partitioned execution, and edge- or fog-assisted training, and a further query added security and privacy terms. We complemented the database search with searches of other digital libraries, a separate literature review on split federated learning, and backward and forward snowballing from included papers. We screened titles and abstracts against the scope rule above and read the remaining candidates in full. The resulting corpus comprises 29 systems: 10 full-delegation systems and 19 partial-delegation systems. The unit of analysis is a distinct system or materially different deployment mode. Our objective is to systematize the guarantees and limitations arising from delegated execution rather than provide an exhaustive inventory of FL research.

\noindent\textbf{Coding and comparison.} We examine each system through its delegated execution path using the framework developed in Section~\ref{sec:model}. For each audited property, we distinguish the guarantee established by the system from the evidence supporting it, and we code only what the primary publication describes or evaluates. Property support is coded as \emph{Supported}, \emph{Limited}, \emph{Unsupported}, \emph{Not Reported}, or \emph{Not Applicable}. \emph{Limited} denotes partial coverage of the audited property, whereas \emph{Unsupported} indicates that the described workflow does not establish it. We do not interpret \emph{Not Reported} as evidence of failure, and we keep it distinct from \emph{Unsupported} because the two convey different information about the reviewed publications. Similarly, the deployment indicators characterize the available resource measurements rather than rank systems according to their intended deployment settings. Appendix~\ref{app:corpus} records the rationale and source locations behind each system's codes. The resulting profiles allow us to examine how guarantees relate across the delegated transition. Our synthesis concerns recurring patterns across the included systems rather than estimates of their prevalence across the wider FL literature.

\section{Delegated-Transition Model}
\label{sec:model}

Delegated FL separates the owner of client data from the party performing some or all computation that determines an owner-attributed contribution. We represent this separation as a stateful delegated transition that serves as a \emph{conceptual and operational audit framework}, not as a cryptographic security definition. The framework identifies what a system exposes, establishes, assumes, and leaves unsupported; formal privacy and integrity claims retain only the guarantees of their underlying mechanisms.

\subsection{Delegated Transition}
\label{sec:delegated-transition}

Let $i\in\mathcal{I}$ index data owners and $t\in\mathcal{T}$ index training rounds. For each owner $i$ and round $t$, let $\mathcal{J}_{i,t}$ denote the set of selected executor indices and define
\[
\mathcal{E}_{i,t}
:=
\{\mathcal{E}_j\in\mathcal{E}\mid j\in\mathcal{J}_{i,t}\}
\]
as the corresponding executor set. Let $\mathcal{O}_i$ denote owner $i$, $\mathcal{A}$ the admission/aggregation role, and $\mathcal{V}$ any verifier. An optional authority $\mathcal{K}$ manages registration, keys, setup material, or attestation roots. These roles remain logically distinct even when co-located or administered by the same organization.

The owner issues a task commitment:
\[
\tau_{i,t}
=
(i,t,U_{i,t},\mathsf{ctx}_t,\mathsf{pol}_{i,t},
\mathcal{E}_{i,t},\mathsf{auth}_{i,t}),
\]
where $U_{i,t}$ denotes the authorized input, commitment, selection policy, or data handle; $\mathsf{ctx}_t$ identifies the relevant model and protocol state; and $\mathsf{pol}_{i,t}$ specifies the permitted learning procedure. The policy may define randomness, numerical representation and tolerance, minibatch or epoch rules, stopping conditions, retries, and other admissible approximations rather than requiring deterministic replay of one nominal computation. $\mathsf{auth}_{i,t}$ records authorization, including its issuer, scope, validity interval, and revocation status where the design supports these properties.

For each selected executor $j\in\mathcal{J}_{i,t}$, we represent the delegated transition operationally as
\begin{equation}
\begin{aligned}
\mathsf{Del}_{i,j,t}:\;&
(\tau_{i,t},s^O_{i,t},s^{E_j}_{i,t})
\longrightarrow
(r_{i,j,t},\pi_{i,j,t},\mathsf{obs}^{E_j}_{i,t}),\\
\mathbf{r}_{i,t}
&:= (r_{i,j,t})_{j\in\mathcal{J}_{i,t}},\\
\boldsymbol{\pi}_{i,t}
&:= (\pi_{i,j,t})_{j\in\mathcal{J}_{i,t}},\\
\Delta_{i,t}
&= g_i(\mathbf{r}_{i,t},s^O_{i,t}),\\
v_{i,t}
&= \mathsf{Verify}(
\tau_{i,t},
\mathbf{r}_{i,t},
\Delta_{i,t},
\boldsymbol{\pi}_{i,t},
s^V_t),\\
a_{i,t}
&= \mathsf{Admit}(\Delta_{i,t},v_{i,t}).
\end{aligned}
\label{eq:delegated-transition}
\end{equation}

Here, $\mathsf{obs}^{E_j}_{i,t}$ contains the artifacts and protocol events visible to executor $\mathcal{E}_j$ while serving owner $i$ in round $t$. Under full delegation, $g_i$ combines or selects the returned executor results without performing additional contribution-producing learning; when a single executor returns the final contribution, $g_i$ may be the identity. Under partial delegation, $g_i$ includes the owner-retained completion step. A specified verifier evaluates $\mathsf{Verify}$ and returns $v_{i,t}$, after which the admission role records the consequential decision $a_{i,t}$, such as acceptance, rejection, weighting, quarantine, or retry. When no verifier is used, $v_{i,t}$ denotes the protocol's default admission evidence.

The representation accommodates adaptive executor selection, cut placement, failures, and conditional execution modes. The executor set $\mathcal{E}_{i,t}$ and the owner-retained completion step may change across rounds as device resources or network conditions change. We therefore code adaptive systems according to their evaluated execution modes. When evidence covers only part of the supported modes, we record the corresponding property as \emph{Limited}. A contribution may be attributed to the owner, executor, both, or a pseudonymous task identifier. For each audited system, we record the concrete binding among administrative identity, protocol identity, and the signing, authentication, attestation, or verification keys used by each role. Full and partial delegation describe execution placement only; neither category implies privacy, correctness, attribution, or feasibility.

\subsection{Accumulated View}
\label{sec:accumulated-view}

For protected secret $S$, observer or coalition $\mathcal{C}$, and task and round horizon $H$, we audit privacy over the joint structured view:
\begin{equation}
\begin{aligned}
&\operatorname{proj}_{\mathcal{C}}(\mathsf{obs}_{i,t})
:= \bigl(\mathsf{obs}^{R}_{i,t}\bigr)_{R\in \mathcal{C}},\\
&\mathsf{View}^{H}_{\mathcal{C}}
:= \Bigl(
\bigl(\operatorname{proj}_{\mathcal{C}}(\mathsf{obs}_{i,t})\bigr)_{(i,t)\in H},
\mathsf{State}^{H}_{\mathcal{C}},\mathsf{Rand}_{\mathcal{C}},\\
&\qquad \qquad \quad
\mathsf{Keys}_{\mathcal{C}},\mathsf{Aux}_{\mathcal{C}}
\Bigr).
\end{aligned}
\label{eq:accumulated-view}
\end{equation}

The tuple represents one ordered, correlated view rather than a set union. For colluding roles, it includes their combined observations, shared state, randomness, keys, auxiliary information, and cross-round correlations. Changes in cut placement or executor assignment may alter the artifacts available to each observer. Consequently, privacy established for one execution mode does not necessarily extend to the accumulated view across different modes. Coalition-specific trust and non-collusion assumptions are recorded explicitly. We code executor--aggregator collusion as \emph{Not Applicable} when no distinct executor--aggregator pair exists and as \emph{Not Reported} when such a pair exists but its collusion assumptions are not reported.

Each privacy row identifies the secret and owner, observer or coalition, horizon, visible artifacts, auxiliary data, allowed leakage $L(S)$, and evidence type. Formal privacy guarantees are recorded only for their stated adversary and leakage function. Empirical attack resistance is recorded separately with the attack, success metric, baseline, and uncovered components of the accumulated view. Failure of a tested attack is not treated as proof of privacy. Appendix~\ref{app:corpus} therefore reports \emph{observable residual view} separately from demonstrated or explicitly permitted leakage.

\subsection{Delegated Attribution}
\label{sec:delegated-attribution}

Delegated attribution is audited through distinct predicates rather than one compound chain. For task $\tau_{i,t}$, we consider
\begin{equation}
\begin{aligned}
\mathcal{P}_{i,t}=\bigl\{&
P_{\mathsf{auth}},P_{\mathsf{input}},P_{\mathsf{state}},
P_{\mathsf{alg}},P_{\mathsf{work}},\\
&
P_{\mathsf{result}},P_{\mathsf{exec}},P_{\mathsf{fresh}},
P_{\mathsf{admit}}
\bigr\}.
\end{aligned}
\label{eq:attribution-predicates}
\end{equation}

These predicates represent owner authorization/revocation, input binding, current model/optimizer-state binding, algorithm and parameter binding, policy-relative amount of work, result consistency, executor identity/accountability, non-equivocation or replay resistance, and consequential admission. For each predicate, the audit records who establishes it, who verifies it, the mechanism used, and the resulting protocol consequence.

$P_{\mathsf{work}}$ is evaluated relative to $\mathsf{pol}_{i,t}$ rather than an implicit requirement to process every record or execute an exact number of deterministic operations. A legitimate policy may permit probabilistic sampling, approximate arithmetic, early stopping, compression, or variable local steps. Assurance may therefore establish compliance exactly, probabilistically, or within a stated tolerance.

The predicates also distinguish owner attribution from executor accountability. Owner authorization can establish that an executor was permitted to act for $\mathcal{O}_i$, while $P_{\mathsf{exec}}$ identifies the party responsible for the resulting execution. Owner attribution does not by itself establish that the owner performed, controlled, or is accountable for malicious delegate behavior.

The main audit groups six of the nine predicates into three comparative indicators. Freshness (Fr) concerns $P_{\mathsf{state}}$ and $P_{\mathsf{fresh}}$, while prescribed-work assurance (W) concerns $P_{\mathsf{alg}}$, $P_{\mathsf{work}}$, and $P_{\mathsf{result}}$. Consequential admission (AD) captures $P_{\mathsf{admit}}$. Appendix~\ref{app:corpus} provides further evidence concerning owner authorization, input binding, and executor accountability. Support for a grouped indicator does not imply that every associated predicate is established.

Mechanisms compose only when the predicates they establish refer to the same delegated transition and their protocol consequences connect. An owner-authenticated $\tau_{i,t}$ may establish $P_{\mathsf{auth}}$, while executor evidence $\boldsymbol{\pi}_{i,t}$ may establish $P_{\mathsf{alg}}$ or $P_{\mathsf{work}}$. Admission-bound attribution additionally requires the relevant evidence to be tied to the same task, executor, state, and result, and for the verification outcome to affect $\mathcal{A}$. The framework records this relation-by-relation coverage; it does not itself constitute a proof system.

\subsection{Constrained-Owner Feasibility}
\label{sec:client-enforceability}
\label{sec:framework}

We represent the complete owner-side workload as an accounting vector:
\begin{equation}
\begin{aligned}
&\mathbf{C}_{i,t}
:= \bigl(W_{i,t},M_{i,t},J_{i,t},B_{i,t},S_{i,t},L_{i,t}\bigr),\\
&\mathbf{C}_{i,1:T}
:= \mathbf{C}^{\mathsf{setup}}_i \oplus
\bigoplus_{t=1}^{T}\mathbf{C}_{i,t},\\
&\bigoplus_{t=1}^{T}\mathbf{C}_{i,t}
:= \left(
\begin{gathered}
\sum_{t=1}^{T}W_{i,t},
\max_{1\le t\le T}M_{i,t},
\sum_{t=1}^{T}J_{i,t},\\
\sum_{t=1}^{T}B_{i,t},
\max_{1\le t\le T}S_{i,t},
\max_{1\le t\le T}L_{i,t}
\end{gathered}
\right),\\
\mathbf{C}_{i,1:T}
&\preceq_{\mathrm{cw}}\mathbf{R}_{i,\mathsf{device}}
\Longleftrightarrow
C_{i,k}\le R_{i,k}\ \text{for every matched }k.
\end{aligned}
\label{eq:client-enforceability}
\end{equation}

Here $W_{i,t}$ is computational work, $M_{i,t}$ peak memory, $J_{i,t}$ energy consumption, $B_{i,t}$ communication volume, $S_{i,t}$ persistent state or storage, and $L_{i,t}$ latency. The vector $\mathbf{R}_{i,\mathsf{device}}$ denotes the corresponding device, network, and deadline budgets, and $\preceq_{\mathrm{cw}}$ denotes component-wise comparison. Each component of $\mathbf{C}_{i,1:T}$ is expressed as a resource demand or cost and compared with its corresponding budget. Recovery includes applicable reconnection, replay-state persistence, key recovery, interrupted rounds, and evidence retransmission. For adaptive delegation, the accounting combines the costs incurred under each execution mode. Average round costs alone are insufficient to establish feasibility when peak resource demands determine whether the owner can participate.

Delegation enables otherwise infeasible participation when, under a matched workload and security requirements,
\[
\begin{aligned}
\mathbf{C}^{\mathsf{local}}_{i,1:T}
&\npreceq_{\mathrm{cw}}\mathbf{R}_{i,\mathsf{device}},\\
\mathbf{C}^{\mathsf{del}}_{i,1:T}
&\preceq_{\mathrm{cw}}\mathbf{R}_{i,\mathsf{device}}.
\end{aligned}
\]
If both workflows satisfy the owner budget, delegation may reduce resource demands but is not necessary for participation under that budget. Conversely, removing local learning does not enable participation if the remaining delegated workflow still exceeds the owner budget.

The component-wise comparison is an accounting approximation rather than an independence assumption. Communication energy depends on radio behavior and duration, memory on batching, verification may overlap communication, and computation may depend on thermal or power limits. Feasibility is therefore asserted only for a specified device class, workload, network, round structure, and matched security configuration. A generic ``IoT'' or ``Raspberry Pi'' label is insufficient; each evaluation must state the actual device budget or report enough measurements to establish it.
\section{Corpus Landscape}
\label{sec:landscape}

The corpus reveals recurring \emph{assurance profiles} rather than a progression from less to more delegation. Full delegation externalizes all semantically necessary learning, whereas partial delegation divides the contribution-producing computation between owner and executor. Table~\ref{tab:corpus-audit-matrix} maps how the audited systems address privacy, attribution, owner burden, and supporting evidence under these two execution modes. The systems range from designs centered primarily on relocating computation to designs that add protection for selected artifacts or provide assurance over parts of the delegated transition. We do not collapse these properties into a single score because they represent different guarantees over different parts of the workflow. Furthermore, the absence of a property outside a system's intended scope does not imply that the system fails to meet its original objectives.

\providecommand{\AuditYes}{\textbf{S}} 
\providecommand{\AuditNo}{\textbf{N}} 
\providecommand{\AuditLimited}{\textbf{L}} 
\providecommand{\AuditNR}{\textbf{NR}} 
\providecommand{\AuditNA}{\textbf{NA}}

\paragraph{Finding derivation.}
The findings in Section~\ref{sec:findings} are cross-system logical patterns over the per-system audit records, not prevalence estimates. We apply the corresponding operational conditions to every system or materially distinct mode in Table~\ref{tab:corpus-audit-matrix}: \AuditYes{} provides positive support, \AuditLimited{} supports only an explicitly partial condition, and \AuditNR{} or \AuditNA{} is never treated as counterevidence. We report a pattern as recurring only when it occurs in multiple systems or modes. The controlled studies in Section~\ref{sec:costs} illustrate these patterns but are not counted as corpus support.

\begin{table*}[!t]
\centering
\caption{Per-system audit of full and partial delegation.
Indicators: \textbf{DA}, executor/host threat model; \textbf{C}, executor--aggregator collusion;
\textbf{L}, labels hidden from or retained away from the executor; \textbf{CG}, formal cryptographic confidentiality;
\textbf{DP}, formal differential privacy; \textbf{IA}, named inference attack; \textbf{XR}, linked/repeated/cross-round privacy analysis;
\textbf{Fr}, freshness or current-state/task binding; \textbf{W}, policy-relative prescribed-work assurance;
\textbf{AD}, verification affects admission; \textbf{LW}, owner retains necessary learning;
\textbf{BX}, returned intermediate locally combined; \textbf{AW}, additional owner-side privacy/integrity work;
\textbf{IoT}, complete constrained-IoT workflow; \textbf{M+C}, peak owner memory and communication reported.
Cell codes: \textbf{S}, supported; \textbf{N}, unsupported; \textbf{L}, limited; \textbf{NR}, not reported; \textbf{NA}, structurally not applicable.}
\label{tab:corpus-audit-matrix}

\scriptsize
\setlength{\tabcolsep}{2.2pt}
\renewcommand{\arraystretch}{1.04}

\begin{adjustbox}{width=\textwidth}
\begin{tabular}{@{}>{\raggedright\arraybackslash}p{3.15cm}
 *{15}{>{\centering\arraybackslash}p{0.78cm}}@{}}
\toprule
& \multicolumn{2}{c}{\textbf{Threat}}
& \multicolumn{5}{c}{\textbf{Privacy}}
& \multicolumn{3}{c}{\textbf{Attribution}}
& \multicolumn{3}{c}{\textbf{Owner burden}}
& \multicolumn{2}{c}{\textbf{Evidence}} \\
\cmidrule(lr){2-3}
\cmidrule(lr){4-8}
\cmidrule(lr){9-11}
\cmidrule(lr){12-14}
\cmidrule(lr){15-16}
\textbf{System}
& \textbf{DA}
& \textbf{C}
& \textbf{L}
& \textbf{CG}
& \textbf{DP}
& \textbf{IA}
& \textbf{XR}
& \textbf{Fr}
& \textbf{W}
& \textbf{AD}
& \textbf{LW}
& \textbf{BX}
& \textbf{AW}
& \textbf{IoT}
& \textbf{M+C} \\
\midrule

\multicolumn{16}{@{}l}{\textit{Full delegation}} \\

\textit{TEE edge offloading}~\cite{dong2020offloading}
& \AuditYes{} & \AuditNR{}
& \AuditYes{} & \AuditNo{} & \AuditNo{} & \AuditNo{} & \AuditNo{}
& \AuditNR{} & \AuditNR{} & \AuditNR{}
& \AuditNo{} & \AuditNo{} & \AuditYes{}
& \AuditNo{} & \AuditNo{} \\

\textit{Smart-contract-enabled OF-RAN}~\cite{jijin2022smart}
& \AuditNR{} & \AuditNR{}
& \AuditNo{} & \AuditNo{} & \AuditNo{} & \AuditNo{} & \AuditNo{}
& \AuditNR{} & \AuditNR{} & \AuditLimited{}
& \AuditNo{} & \AuditNo{} & \AuditNo{}
& \AuditNo{} & \AuditNo{} \\

\textit{Zkfhed}~\cite{zhang2025zkfhed}
& \AuditYes{} & \AuditNR{}
& \AuditYes{} & \AuditYes{} & \AuditNo{} & \AuditNo{} & \AuditNo{}
& \AuditLimited{} & \AuditYes{} & \AuditYes{}
& \AuditNo{} & \AuditYes{} & \AuditYes{}
& \AuditNo{} & \AuditNo{} \\

\textit{OVeriFL}~\cite{han2025outsourced}
& \AuditYes{} & \AuditNR{}
& \AuditNo{} & \AuditNo{} & \AuditYes{} & \AuditLimited{} & \AuditLimited{}
& \AuditNo{} & \AuditNo{} & \AuditYes{}
& \AuditNo{} & \AuditYes{} & \AuditYes{}
& \AuditNo{} & \AuditNo{} \\

\textit{TD3 node selection}~\cite{lin2022td3}
& \AuditNR{} & \AuditNR{}
& \AuditNo{} & \AuditNo{} & \AuditNo{} & \AuditNo{} & \AuditNo{}
& \AuditNR{} & \AuditNR{} & \AuditNR{}
& \AuditNo{} & \AuditNo{} & \AuditNo{}
& \AuditNo{} & \AuditNo{} \\

\textit{WW-FL}~\cite{marx2023ww}
& \AuditYes{} & \AuditYes{}
& \AuditYes{} & \AuditYes{} & \AuditNo{} & \AuditNo{} & \AuditNo{}
& \AuditYes{} & \AuditYes{} & \AuditYes{}
& \AuditNo{} & \AuditNo{} & \AuditYes{}
& \AuditNo{} & \AuditNo{} \\

\textit{FedKeeper}~\cite{chadha2020towards}
& \AuditNR{} & \AuditNR{}
& \AuditNo{} & \AuditNo{} & \AuditNo{} & \AuditNo{} & \AuditNo{}
& \AuditLimited{} & \AuditNo{} & \AuditYes{}
& \AuditNo{} & \AuditNo{} & \AuditNo{}
& \AuditYes{} & \AuditNo{} \\

\textit{Quantum private distributed learning}~\cite{li2021quantum}
& \AuditYes{} & \AuditNR{}
& \AuditYes{} & \AuditYes{} & \AuditLimited{} & \AuditYes{} & \AuditNo{}
& \AuditNo{} & \AuditLimited{} & \AuditNo{}
& \AuditNo{} & \AuditYes{} & \AuditYes{}
& \AuditNo{} & \AuditNo{} \\

\textit{FedLess}~\cite{grafberger2021fedless}
& \AuditNR{} & \AuditNR{}
& \AuditNo{} & \AuditNo{} & \AuditLimited{} & \AuditNo{} & \AuditYes{}
& \AuditLimited{} & \AuditNo{} & \AuditLimited{}
& \AuditNo{} & \AuditNo{} & \AuditNo{}
& \AuditNo{} & \AuditNo{} \\

\textit{FedLesScan}~\cite{elzohairy2022fedlesscan}
& \AuditLimited{} & \AuditNR{}
& \AuditNo{} & \AuditNo{} & \AuditNo{} & \AuditNo{} & \AuditNo{}
& \AuditYes{} & \AuditNo{} & \AuditYes{}
& \AuditNo{} & \AuditNo{} & \AuditNo{}
& \AuditNo{} & \AuditNo{} \\

\midrule
\multicolumn{16}{@{}l}{\textit{Partial delegation}} \\

\textit{SGBoost+}~\cite{xu2025sgboost}
& \AuditYes{} & \AuditNo{}
& \AuditYes{} & \AuditYes{} & \AuditNo{} & \AuditYes{} & \AuditLimited{}
& \AuditNo{} & \AuditNo{} & \AuditNo{}
& \AuditYes{} & \AuditYes{} & \AuditYes{}
& \AuditNo{} & \AuditNo{} \\

\textit{SplitFed Learning}~\cite{thapa2022splitfed}
& \AuditYes{} & \AuditNR{}
& \AuditNo{} & \AuditNo{} & \AuditYes{} & \AuditNo{} & \AuditYes{}
& \AuditNo{} & \AuditNo{} & \AuditNo{}
& \AuditYes{} & \AuditYes{} & \AuditYes{}
& \AuditNo{} & \AuditNo{} \\

\textit{Vehicular FSL paradigm}~\cite{wu2023federated}
& \AuditYes{} & \AuditNo{}
& \AuditNo{} & \AuditNo{} & \AuditYes{} & \AuditYes{} & \AuditYes{}
& \AuditNo{} & \AuditNo{} & \AuditNo{}
& \AuditYes{} & \AuditYes{} & \AuditYes{}
& \AuditNo{} & \AuditNo{} \\

\textit{HierSFL}~\cite{quan2023hiersfl}
& \AuditNR{} & \AuditNR{}
& \AuditNo{} & \AuditNo{} & \AuditYes{} & \AuditNo{} & \AuditNo{}
& \AuditNo{} & \AuditNo{} & \AuditNo{}
& \AuditYes{} & \AuditYes{} & \AuditYes{}
& \AuditNo{} & \AuditNo{} \\

\textit{DT-ASFL}~\cite{stephanie2023dtasfl}
& \AuditYes{} & \AuditNA{}
& \AuditNo{} & \AuditYes{} & \AuditNo{} & \AuditYes{} & \AuditNo{}
& \AuditYes{} & \AuditNo{} & \AuditYes{}
& \AuditYes{} & \AuditYes{} & \AuditYes{}
& \AuditNo{} & \AuditNo{} \\

\textit{PPTADI}~\cite{meng2025pptadi}
& \AuditYes{} & \AuditNR{}
& \AuditYes{} & \AuditNo{} & \AuditNo{} & \AuditYes{} & \AuditNo{}
& \AuditNR{} & \AuditNR{} & \AuditNR{}
& \AuditYes{} & \AuditYes{} & \AuditYes{}
& \AuditNo{} & \AuditNo{} \\

\textit{APP-SplitFed}~\cite{stephanie2025weight}
& \AuditNR{} & \AuditNA{}
& \AuditNo{} & \AuditYes{} & \AuditNo{} & \AuditYes{} & \AuditNo{}
& \AuditYes{} & \AuditNo{} & \AuditYes{}
& \AuditYes{} & \AuditYes{} & \AuditYes{}
& \AuditNo{} & \AuditNo{} \\

\textit{USFL}~\cite{zhao2023usfl}
& \AuditYes{} & \AuditNA{}
& \AuditYes{} & \AuditNo{} & \AuditNo{} & \AuditNo{} & \AuditNo{}
& \AuditNR{} & \AuditNR{} & \AuditNR{}
& \AuditYes{} & \AuditYes{} & \AuditNo{}
& \AuditNo{} & \AuditNo{} \\

\textit{Privacy-aware LLM SFL}~\cite{chen2025llmiot}
& \AuditLimited{} & \AuditNA{}
& \AuditNR{} & \AuditNo{} & \AuditNo{} & \AuditLimited{} & \AuditNo{}
& \AuditNR{} & \AuditNR{} & \AuditNR{}
& \AuditYes{} & \AuditYes{} & \AuditNo{}
& \AuditNo{} & \AuditNo{} \\

\textit{DP-MORA}~\cite{huang2025dpmora}
& \AuditYes{} & \AuditNA{}
& \AuditNo{} & \AuditNo{} & \AuditNo{} & \AuditYes{} & \AuditNo{}
& \AuditNR{} & \AuditNR{} & \AuditNR{}
& \AuditYes{} & \AuditYes{} & \AuditYes{}
& \AuditYes{} & \AuditNo{} \\

\textit{ShieldTSE}~\cite{chen2024shieldtse}
& \AuditYes{} & \AuditNA{}
& \AuditYes{} & \AuditNo{} & \AuditNo{} & \AuditYes{} & \AuditNo{}
& \AuditNR{} & \AuditNR{} & \AuditNR{}
& \AuditYes{} & \AuditYes{} & \AuditYes{}
& \AuditNo{} & \AuditNo{} \\

\textit{VerifyVFL}~\cite{hou2025verifyvfl}
& \AuditYes{} & \AuditNA{}
& \AuditYes{} & \AuditYes{} & \AuditNo{} & \AuditYes{} & \AuditNo{}
& \AuditNo{} & \AuditYes{} & \AuditYes{}
& \AuditYes{} & \AuditYes{} & \AuditYes{}
& \AuditNo{} & \AuditNo{} \\

\textit{SFL with HMAC}~\cite{latif2025hmac}
& \AuditNR{} & \AuditNR{}
& \AuditNR{} & \AuditNo{} & \AuditNo{} & \AuditNo{} & \AuditNo{}
& \AuditNo{} & \AuditNo{} & \AuditYes{}
& \AuditYes{} & \AuditYes{} & \AuditYes{}
& \AuditNo{} & \AuditNo{} \\

\textit{SFsL}~\cite{okegbile2025novel}
& \AuditNR{} & \AuditNA{}
& \AuditNR{} & \AuditNo{} & \AuditNo{} & \AuditNo{} & \AuditNo{}
& \AuditNR{} & \AuditNR{} & \AuditYes{}
& \AuditYes{} & \AuditYes{} & \AuditYes{}
& \AuditNo{} & \AuditNo{} \\

\textit{Aergia}~\cite{cox2022aergia}
& \AuditNR{} & \AuditNR{}
& \AuditYes{} & \AuditNo{} & \AuditNo{} & \AuditNo{} & \AuditNo{}
& \AuditYes{} & \AuditNo{} & \AuditNo{}
& \AuditYes{} & \AuditNo{} & \AuditYes{}
& \AuditNo{} & \AuditNo{} \\

\textit{HSFL}~\cite{xia2022hsfl}
& \AuditNR{} & \AuditNR{}
& \AuditNo{} & \AuditNo{} & \AuditNo{} & \AuditNo{} & \AuditNo{}
& \AuditNR{} & \AuditNR{} & \AuditNR{}
& \AuditYes{} & \AuditYes{} & \AuditNo{}
& \AuditNo{} & \AuditNo{} \\

\textit{HFSL}~\cite{wu2025hfsl}
& \AuditNR{} & \AuditNR{}
& \AuditNo{} & \AuditNo{} & \AuditNo{} & \AuditNo{} & \AuditNo{}
& \AuditNR{} & \AuditNR{} & \AuditNR{}
& \AuditYes{} & \AuditYes{} & \AuditNo{}
& \AuditNo{} & \AuditNo{} \\

\textit{DRWB}~\cite{fei2025privacy}
& \AuditNo{} & \AuditNo{}
& \AuditNo{} & \AuditNo{} & \AuditNo{} & \AuditNo{} & \AuditNo{}
& \AuditNo{} & \AuditNo{} & \AuditNo{}
& \AuditYes{} & \AuditNo{} & \AuditNo{}
& \AuditNo{} & \AuditNo{} \\

\textit{DT-MEC-assisted FL}~\cite{he2023computation}
& \AuditNo{} & \AuditNo{}
& \AuditNo{} & \AuditNo{} & \AuditNo{} & \AuditNo{} & \AuditNo{}
& \AuditNo{} & \AuditNo{} & \AuditNo{}
& \AuditYes{} & \AuditNo{} & \AuditNo{}
& \AuditNo{} & \AuditNo{} \\

\bottomrule
\end{tabular}
\end{adjustbox}
\end{table*}

\subsection{Mode-Level Patterns}
\label{sec:mode-patterns}

Full-delegation systems vary considerably in the assurance mechanisms surrounding externalized learning. WW-FL~\cite{marx2023ww} and Zkfhed~\cite{zhang2025zkfhed}, for example, combine delegated execution with explicit privacy and attribution mechanisms, while several other systems concentrate primarily on establishing the execution or orchestration path. The corpus therefore distinguishes systems not by the amount of computation moved, which is already complete under this mode, but by the protections and evidence attached to that external execution.

Partial delegation presents a broader range of boundary designs. SGBoost+~\cite{xu2025sgboost}, SplitFed Learning~\cite{thapa2022splitfed}, the vehicular FSL paradigm~\cite{wu2023federated}, PPTADI~\cite{meng2025pptadi}, DP-MORA~\cite{huang2025dpmora}, ShieldTSE~\cite{chen2024shieldtse}, and VerifyVFL~\cite{hou2025verifyvfl} protect or evaluate different representations exchanged between owner and executor. These systems also differ in how much learning remains with the owner and whether additional privacy or integrity work is required locally. The resulting profiles depend on the placement of the learning boundary and on the artifacts exchanged across it, rather than on a simple degree of offloading.

The motivation for delegation also differs across the corpus. In serverless systems such as FedLess~\cite{grafberger2021fedless} and FedLesScan~\cite{elzohairy2022fedlesscan}, external execution is motivated by resource allocation and scalability. In vertical FL systems such as SGBoost+ and VerifyVFL, the delegated computation arises from the distribution of features and learning operations across parties. We include these systems because their execution paths instantiate the owner--executor separation, not because they necessarily target owners that cannot train locally.

The attribution columns show similarly distinct profiles. WW-FL~\cite{marx2023ww} supports all three audited indicators of freshness, prescribed-work assurance, and consequential admission. FedLesScan~\cite{elzohairy2022fedlesscan}, DT-ASFL~\cite{stephanie2023dtasfl}, and APP-SplitFed~\cite{stephanie2025weight} support freshness and admission without prescribed-work assurance, while VerifyVFL~\cite{hou2025verifyvfl} supports prescribed-work assurance and admission without corresponding freshness support. Several other systems provide admission-related checks without establishing the work that produced the accepted result. These combinations show why the three indicators are coded separately rather than treated as a single integrity property.

\subsection{Cost Relocation and Constrained-Owner Feasibility}
\label{sec:cost-relocation}

The owner-burden profiles differ substantially across the two execution modes. In full delegation, TEE edge offloading and Zkfhed remove necessary owner-side training but retain input preparation or protection, communication, setup or key state, and attestation or verification work~\cite{dong2020offloading,zhang2025zkfhed}. In partial delegation, SplitFed Learning, HierSFL, and VerifyVFL retain necessary owner-side learning, returned intermediate processing, and additional privacy or integrity work, whereas USFL retains the split-completion workload without the same coded additional assurance burden~\cite{thapa2022splitfed,quan2023hiersfl,hou2025verifyvfl,zhao2023usfl}. Aergia presents a different profile in which local learning and assurance-related work remain without owner-local recombination of a returned intermediate~\cite{cox2022aergia}.

Evidence for these residual workloads is considerably less complete. FedKeeper~\cite{chadha2020towards} and DP-MORA~\cite{huang2025dpmora} provide constrained-IoT deployment evidence, while much of the corpus relies on simulation, server-class prototypes, or incomplete owner-side measurements. No audited system jointly reports peak owner memory and communication. This measurement gap also occurs in explicitly IoT-oriented systems, including SFL with HMAC~\cite{latif2025hmac}, PPTADI~\cite{meng2025pptadi}, and DT-ASFL~\cite{stephanie2023dtasfl}. The absence of such measurements does not establish that a system is infeasible. Rather, it limits what can be concluded about its suitability for a constrained owner. The corpus therefore supports comparison of specific cost-relocation patterns, but provides limited evidence for complete constrained-owner feasibility.
\section{Cross-Axis Findings}
\label{sec:findings}

The principal findings arise from intersections among privacy, delegated attribution, and constrained-owner feasibility rather than from marginal indicator counts. We derive them as recurring patterns over the per-system audit records. The findings remain bounded to the reviewed systems and the conservative coding rules in Section~\ref{sec:method}.

\subsection{Finding 1: Privacy Is Determined by the Accumulated View}

The corpus shows that privacy mechanisms protect different portions of the information available to the executor. OVeriFL protects the released contribution while its delegated trainer receives upstream training information~\cite{han2025outsourced}; U-shaped FL retains labels and loss computation locally while exposing intermediate representations and their gradients to the remote segment~\cite{zhao2023usfl}; and SGBoost+ and PPTADI protect or evaluate different artifacts produced during split execution~\cite{xu2025sgboost,meng2025pptadi}. These mechanisms therefore provide different forms of coverage rather than interchangeable privacy guarantees.

The resulting privacy boundary is both artifact-specific and temporal. Federated training repeatedly exposes the executor to observations generated for the same owner under evolving model and protocol states. When these observations are linkable, the executor may correlate information across rounds to infer properties that are not apparent from an isolated release. Furthermore, in adaptive delegation, changes in cut placement or executor assignment may expose different artifacts over time. Protecting a single per-round artifact therefore does not establish privacy of the accumulated delegated transcript. This does not imply that leakage increases monotonically with the number of rounds. The effect depends on the protected secret, the observable artifacts, and how these observations are linked across the model trajectory.

\subsection{Finding 2: Privacy Evidence Does Not Automatically Compose}

The corpus also differs in the type and scope of evidence used to support privacy claims. Quantum private distributed learning combines formal cryptographic protection with a named inference evaluation, whereas ShieldTSE and DP-MORA provide empirical inference analysis without a corresponding formal cryptographic or differential-privacy guarantee under our coding~\cite{li2021quantum,chen2024shieldtse,huang2025dpmora}. Privacy-aware LLM SFL provides a further representation-oriented evaluation with only limited attack evidence in the audited dimensions~\cite{chen2025llmiot}. The distinction is therefore not simply how much evidence a system provides, but what privacy claim that evidence actually establishes.

Privacy evidence supports only the claim established under its stated assumptions. A formal guarantee over one transcript does not necessarily extend to other representations exchanged during delegated learning. Similarly, empirical resistance to an inference attack does not establish privacy against a different observer or over a longer execution horizon. Such results strengthen a common privacy claim only when their scopes align. Otherwise, each result may remain valid while other parts of the executor's accumulated view remain uncharacterized.

\subsection{Finding 3: Assurance Does Not Propagate Across the Delegated Path}

The attribution audit reveals recurring partial bindings across the delegated transition. WW-FL is the only system in our audit coded as supporting freshness, prescribed-work assurance, and consequential admission~\cite{marx2023ww}. FedLesScan, DT-ASFL, and APP-SplitFed support freshness and admission without corresponding prescribed-work assurance~\cite{elzohairy2022fedlesscan,stephanie2023dtasfl,stephanie2025weight}; VerifyVFL supports prescribed-work assurance and admission without the same freshness relation~\cite{hou2025verifyvfl}; and HMAC-based SplitFed provides admission-related checking without establishing either relation under the narrow coding used here~\cite{latif2025hmac}. The recurring pattern is therefore not simply different levels of verification, but assurance over different links in the owner-to-admission path.

Assurance over one link does not establish the adjacent ones. Freshness can bind a result to a current task without showing that the prescribed work was performed. Conversely, computation evidence can establish a work relation without establishing freshness. Authentication or admission checks can reject malformed or unauthorized messages without establishing how an accepted result was produced. A delegated result is therefore only as attributable as the relations established and connected across the delegated transition. System-level claims such as ``verified delegation'' should consequently state which relations are established rather than treating verification as a binary property.

\subsection{Finding 4: Boundary Placement Redistributes Owner Burden and Assurance}

The owner--executor boundary determines not only where computation occurs, but also which artifacts cross between roles and which assurance tasks remain local. SplitFed Learning and HierSFL retain necessary owner-side learning, require a returned intermediate to be combined locally, and add protection- or integrity-related owner work~\cite{thapa2022splitfed,quan2023hiersfl}. USFL retains learning and the completion boundary without the same coded additional owner work~\cite{zhao2023usfl}, while Aergia exhibits a different combination of retained learning and additional assurance burden~\cite{cox2022aergia}. These systems therefore relocate different parts of the owner workflow even when they all reduce some portion of local execution.

The corpus shows that greater computational offloading does not imply a proportional reduction in owner burden. Moving the boundary can reduce local learning while increasing communication or introducing additional protection and verification work. It also changes which artifacts require protection and where assurance must be established. For adaptive delegation, these obligations may vary across rounds. A configuration that reduces average computational cost may still exceed the owner's peak resource budget. Boundary placement is therefore a joint security-and-resource decision: two designs that offload similar amounts of computation can leave materially different residual obligations and trust assumptions for the owner.

\subsection{Finding 5: Assurance Coverage and Deployment Evidence Are Misaligned}

Deployment evidence does not consistently track the privacy or attribution coverage observed in the corpus. FedKeeper and DP-MORA provide constrained-IoT workflow evidence~\cite{chadha2020towards,huang2025dpmora}, yet neither jointly reports the peak owner-memory and communication evidence required by the narrow M+C indicator. This measurement gap also occurs in other IoT-oriented systems, including HMAC-based SplitFed, PPTADI, and DT-ASFL~\cite{latif2025hmac,meng2025pptadi,stephanie2023dtasfl}. Conversely, several systems with richer privacy or attribution mechanisms do not evaluate the complete protected owner workflow on constrained IoT hardware. The mismatch therefore extends beyond systems whose original objectives do not concern constrained devices.

This mismatch limits what can be concluded about constrained-owner feasibility from either security mechanisms or deployment claims alone. A system may establish substantial privacy or attribution properties without measuring the owner-side cost of realizing them. Similarly, an IoT-oriented evaluation may demonstrate execution feasibility without exercising the complete protected workflow. This does not establish that the corresponding systems are infeasible. Rather, evidence of security coverage and constrained deployment needs to refer to the same configuration and owner path before it jointly supports a secure-and-feasible deployment claim.

\paragraph{Cross-finding synthesis.}
Taken together, the findings reveal a recurring composition gap between mechanism-level guarantees and the complete owner-attributed delegated transition. Privacy protection may cover an individual artifact while other correlated observations remain exposed. Similarly, attribution evidence may establish a particular execution relation without connecting it to the remaining parts of the delegated transition. Even when these relations are established, the owner may still face resource demands that exceed its device budget.

VerifyVFL~\cite{hou2025verifyvfl} illustrates this distinction. The owner delegates gradient computation and verifies the returned gradients before using them in the retained learning procedure. However, the verified work relation does not establish round freshness under our audit. Furthermore, the privacy evaluation does not characterize the accumulated observations available to the executor, and the owner-side learning and verification costs are not jointly evaluated against a constrained-device budget. The gradient-verification guarantee remains meaningful, but it does not by itself establish that the resulting delegated contribution is private, attributable to the current task, and feasible for a constrained owner.

WW-FL~\cite{marx2023ww} provides a different assurance profile. The system externalizes learning while providing cryptographic protection and supporting all three attribution indicators in our audit. However, the feasibility of the complete protected workflow for a constrained owner remains unestablished. Unlike VerifyVFL, the gap identified here concerns the alignment between security coverage and deployment evidence rather than the absence of a particular attribution indicator.

These examples distinguish incomplete assurance coverage from incomplete deployment evidence. Neither implies that the underlying mechanisms fail to provide their stated guarantees. The composition gap arises when these guarantees and their supporting evidence do not jointly establish the same end-to-end claim. Evaluating them over the delegated transition makes it possible to identify which part of that claim remains unestablished.
\section{Controlled Studies and Owner-Side Costs}
\label{sec:costs}
\label{sec:complexity}
\label{sec:complexity_analysis}

Reported overheads in the reviewed systems are not directly comparable because workloads, hardware, protection mechanisms, threat models, and measurement boundaries differ. We therefore use two controlled mechanism-level studies on a fixed deployment. Study~I tests whether encrypted inputs remain protected when a correlated residual is exposed and how reconstruction depends on cross-round linkage. Study~II tests whether local labels and authenticated split messages are sufficient for privacy or prescribed execution. These studies instantiate recurring corpus patterns rather than reproduce individual systems. We use matched configurations to isolate the effects of protection and boundary placement on the same learning workload.

\subsection{Measurement Model and Experimental Design}

For role $r$ and configuration $c$, we record
\begin{equation}
\mathcal{R}_{r,c} =
\left(
T^{\mathsf{active}},
M^{\mathsf{peak}},
B^{\mathsf{tx}},
B^{\mathsf{rx}},
S^{\mathsf{persist}},
T^{\mathsf{cold}},
T^{\mathsf{online}},
T^{\mathsf{e2e}}
\right)_{r,c},
\label{eq:empirical-resource-record}
\end{equation}
denoting active CPU time, peak resident memory, transmitted and received bytes, persistent state, cold-start time, protocol-online time, and end-to-end round latency. Bytes are measured at the socket boundary; energy is omitted because the testbed lacks reproducible direct power measurements.

Under full delegation, owner cost can remain in input protection, key handling, communication, result processing, and verification. Under partial delegation, the owner retains part of the learning path and exchanges boundary activations and gradients. For cut $k$, their application payload over $T$ rounds is
\begin{equation}
B_{\mathsf{payload}}(k)
=
\sum_{t=1}^{T}
\sum_{b=1}^{N_{b,t}}
\left(
s_a |a_{k,t,b}|+
s_g |\nabla a_{k,t,b}|
\right),
\label{eq:split-communication}
\end{equation}
where measured wire volume additionally includes serialization, synchronization, authentication, framing, and control traffic.

\paragraph{Testbed and workload.}
Experiments use five physical hosts on a dedicated local wireless network. $O_1$ and $O_2$ are Raspberry Pi~4 Model~B devices with 4\,GB RAM, $O_3$ is a Raspberry Pi~5 with 8\,GB RAM, $E$ is an Intel Core i9-13900 workstation with 32\,GB RAM, and aggregator $A$ is an Intel Core i5-3470 host with 8\,GB RAM. Roles communicate directly over length-prefixed TCP. Study~I uses Go, Lattigo library~\cite{lattigo}, and CKKS~\cite{cheon2017homomorphic}; only $A$ holds the CKKS secret key. Study~II uses Python with PyTorch, NumPy, and scikit-learn. The artifact records software versions, cryptographic parameters, scripts, attack code, configuration hashes, and outputs~\cite{dfl2026}.

Both studies use PAMAP2~\cite{reiss2012introducing}, excluding optional recordings and activity~0 and retaining 12 activities. Subjects are fixed as $O_1=\{S_1,S_2,S_3\}$, $O_2=\{S_4,S_5,S_6\}$, and $O_3=\{S_7,S_8,S_9\}$. Windows are 2.56\,s with 1.28\,s stride and at least 90\% activity purity. Partitioning occurs before window generation, with guard regions at incompatible boundaries; the resulting 1,476-record manifest has no raw-timestamp overlap between learning and test partitions or between attacker-auxiliary and target partitions.

Learning and privacy experiments use seeds $\{101,211,307\}$. Study~II reports mean$\pm$sample-SD across the three seeds. Study~I uses deterministic full-batch optimization, so its three executions are implementation replications. System measurements use three fresh-process repetitions per configuration and report medians.

\paragraph{Threat models and corpus grounding.}
Table~\ref{tab:controlled-configurations} summarizes secrets, views, keys, linkage, deviations, and trust. I-OR is an I-ER ablation that opens the residual to $E$; citations otherwise ground execution/protection patterns, not exact implementations.

\begin{table}[!t]
\centering
\caption{Corpus grounding and threat models of the controlled configurations.}
\label{tab:controlled-configurations}
\scriptsize
\setlength{\tabcolsep}{1.2pt}
\renewcommand{\arraystretch}{1.02}

\begin{tabularx}{\columnwidth}{@{}
>{\raggedright\arraybackslash}p{0.10\columnwidth}
>{\raggedright\arraybackslash}p{0.20\columnwidth}
>{\raggedright\arraybackslash}p{0.19\columnwidth}
>{\raggedright\arraybackslash}p{0.17\columnwidth}
>{\raggedright\arraybackslash}X
@{}}
\toprule
\textbf{Config.} &
\textbf{Pattern} &
\textbf{Corpus} &
\textbf{Objective / secret} &
\textbf{Adversary view} \\
\midrule

\multicolumn{5}{@{}l}{\textit{Study I --- full delegation}} \\
\cmidrule(r){1-5}

I-L
& Local training
& Baseline
& Utility / cost
& No executor \\

I-P
& Plaintext delegation
& \cite{chadha2020towards,grafberger2021fedless,elzohairy2022fedlesscan}
& Visibility
& $E$: plaintext $x,y$ \\

I-OR
& Protected input, open residual
& I-ER ablation
& $x,y$, activity
& $E$: ciphertext + residual; key at $A$ \\

I-ER
& Protected input and residual
& \cite{zhang2025zkfhed,li2021quantum,marx2023ww}
& $x,y$, activity
& $E$: ciphertext only; key at $A$ \\

\addlinespace[2pt]

\multicolumn{5}{@{}l}{\textit{Study II --- partial delegation}} \\
\cmidrule(r){1-5}

II-L
& Local FL
& Baseline
& Utility / cost
& No executor \\

II-S
& Split partitioning
& \cite{thapa2022splitfed,quan2023hiersfl}
& Split reference
& $E$: split tensors, state \\

II-U-E
& U-shaped, labels local
& \cite{zhao2023usfl,meng2025pptadi}
& Activity, identity
& $E$: $z,g$ \\

II-U-D
& Deeper split
& \cite{zhao2023usfl,huang2025dpmora,chen2025llmiot}
& Activity, identity
& $E$: deeper-cut $z,g$ \\

II-U-H
& Authenticated split
& \cite{latif2025hmac}
& Authenticity
& $E$: messages; HMAC key \\

\bottomrule
\end{tabularx}

\vspace{2pt}
\begin{minipage}{\columnwidth}
\scriptsize
\textit{Deviations.}
Only II-U-H tests active deviations: a no-key attacker may modify, replay, or mismatch messages; a key-holding $E$ may partially execute, reuse stale work, negate gradients, or substitute results.

\textit{Privacy assumptions.}
Study~I varies cross-round linkage; Study~II uses disjoint attacker-auxiliary and target partitions.

\textit{Common assumptions.}
Except in the specified II-U-H deviation tests, $E$ follows the protocol. Owners and $A$ are trusted. Collusion, malicious owners, verifier compromise, denial of service, and unauthorized key compromise are excluded.
\end{minipage}
\end{table}

\subsection{Study I: Protected Input, Exposed Residual}
\label{sec:studies}
\label{sec:evaluation}

Study~I tests whether encrypting owner input is sufficient when the delegated interface returns a correlated plaintext residual. Each record contains the temporal mean of 12 chest-IMU channels; median heart rate is the regression target and activity a secondary attribute. The fixed partitions contain 360 learning, 144 utility-test, 36 attacker-auxiliary, and 108 attacker-target records.

We train $\hat y=w^\top x+b$ with mean-squared error using synchronous full-batch federated gradient descent for 10 rounds. $O_2$ and $O_3$ always train locally. For $O_1$, \textsc{I-L} is local training, \textsc{I-P} plaintext delegation, \textsc{I-OR} encrypted input and target with a residual opened to $E$, and \textsc{I-ER} the matched control in which the residual remains encrypted. In \textsc{I-OR},
\begin{equation}
e_{n,t}=w_t^\top x_n+b_t-y_n,
\label{eq:opened-residual}
\end{equation}
and $A$ decrypts the residual before returning it to $E$. In \textsc{I-ER}, the residual and feature gradient remain encrypted until $A$ decrypts the final local gradient. Operation tracing confirms that $O_1$ performs no semantically necessary learning in either delegated HE configuration.

\paragraph{Residual disclosure and reconstruction.}
Encrypted execution closely matches plaintext: across 60 round-level comparisons, maximum absolute gradient error is $6.22\times10^{-9}$ and the largest final-RMSE difference is $9.24\times10^{-12}$. Because $w_0=b_0=0$,
\begin{equation}
e_{n,1}=-y_n,
\end{equation}
so \textsc{I-OR} directly exposes the heart-rate target in the first round, independently of cross-round linkage, up to CKKS approximation. The measured target MAE is $5.81\times10^{-11}$.

We use feature relative-$L_2$ error throughout the reconstruction analysis. For true target-feature matrix $X$ and reconstruction $\hat X$, we compute
\begin{equation}
E_{\mathrm{rel}\text{-}L_2}
=
\frac{\|\hat X-X\|_F}{\|X\|_F},
\label{eq:feature-relative-l2}
\end{equation}
over the records used in each reconstruction analysis and their 12 features. The original linkage and round-count analyses use the 120 learning records of $O_1$, rather than the 108 attacker-target records across all owners. Lower values indicate better reconstruction. The reported linkage values are means across three executions; figure error bars show their sample SD.

Let $R_{\mathrm{rec}}$ denote the number of linked rounds used for feature reconstruction. Because each record contains 12 unknown input features, the reconstruction systems at $R_{\mathrm{rec}}\in\{1,5,10\}$ have ranks 1, 5, and 10, respectively, and remain underdetermined throughout this sweep. At $R_{\mathrm{rec}}=1$, the reported condition number of 1 characterizes only the single resolved direction and is not evidence of a well-conditioned full reconstruction problem. Under stable linkage, $E_{\mathrm{rel}\text{-}L_2}$ changes from $1.000$ to $0.0817$ to $34.74$ for $R_{\mathrm{rec}}\in\{1,5,10\}$. At $R_{\mathrm{rec}}=5$, the reported condition number is $1.37\times10^{4}$; at $R_{\mathrm{rec}}=10$, it rises to $5.29\times10^{13}$, indicating severe ill-conditioning in the resolved system. The minimum-norm reconstruction recovers the component of the input in the row space of the accumulated equations, not the complete input. With zero initialization, the first residual reveals $y_n$, allowing subsequent residuals to be written as linear equations in $x_n$. A low aggregate relative-$L_2$ error does not imply that every feature is recovered accurately. At five linked rounds, the relative error is 0.0161 for magnetometer features but 0.9906 for gyroscope features. The corresponding five-round analysis on the 36 attacker-target records of $O_1$ yields an aggregate error of 0.0738. Reconstruction is therefore non-monotone and depends on the model trajectory, feature scale, and numerical conditioning of the accumulated equations. In a supplementary reconstruction analysis, the ten-round error is $32.99\pm7.25$, while ridge regularization reduces the error to 0.061. The high unregularized error should therefore not be interpreted as a privacy guarantee.

With nonzero initialization, the first residual no longer directly reveals the target. The supplementary analysis yields a five-round reconstruction error of $0.885\pm0.051$ and a ten-round ridge reconstruction error of $0.874\pm0.059$. Nevertheless, each opened residual still provides a linear relation involving the private features and target. These results distinguish the direct disclosure caused by zero initialization from the more general leakage through repeated residual exposure.

\begin{figure}[t]
\centering
\begin{tikzpicture}
\begin{axis}[
    width=7.5cm, height=4cm,
    ymode=log,
    log basis y=10,
    xlabel={Cross-round linkage},
    ylabel={Feature relative-$L_2$ error},
    symbolic x coords={Stable,75,50,25,Permuted,Shuffled},
    xtick=data,
    xticklabels={
        {Stable\\(100\%)},
        {75\%\\correct},
        {50\%\\correct},
        {25\%\\correct},
        {Permuted\\(0\%)},
        {Shuffled\\(0\%)}
    },
    x tick label style={align=center,font=\scriptsize},
    tick label style={font=\scriptsize},
    label style={font=\small},
    grid=major,
    ymin=0.05,
]
\addplot+[
    mark=*,
    thick,
    error bars/.cd,
    y dir=both,
    y explicit,
]
coordinates {
    (Stable,0.081729) +- (0,0.0000000000033)
    (75,358.008)      +- (0,0.0000002)
    (50,723.522)      +- (0,0.0000004)
    (25,1086.914)     +- (0,0.0000006)
    (Permuted,1306.071)+- (0,0.0000007)
    (Shuffled,1828.379)+- (0,169.296)
};
\end{axis}
\end{tikzpicture}
\caption{Study-I feature reconstruction under varying cross-round linkage at
$R_{\mathrm{rec}}=5$, measured using feature relative-$L_2$ error.
The analysis uses the 120 learning records of $O_1$. Lower is better; the y-axis is logarithmic. Points show mean$\pm$SD
across three executions.}
\label{fig:study1-linkage}
\end{figure}
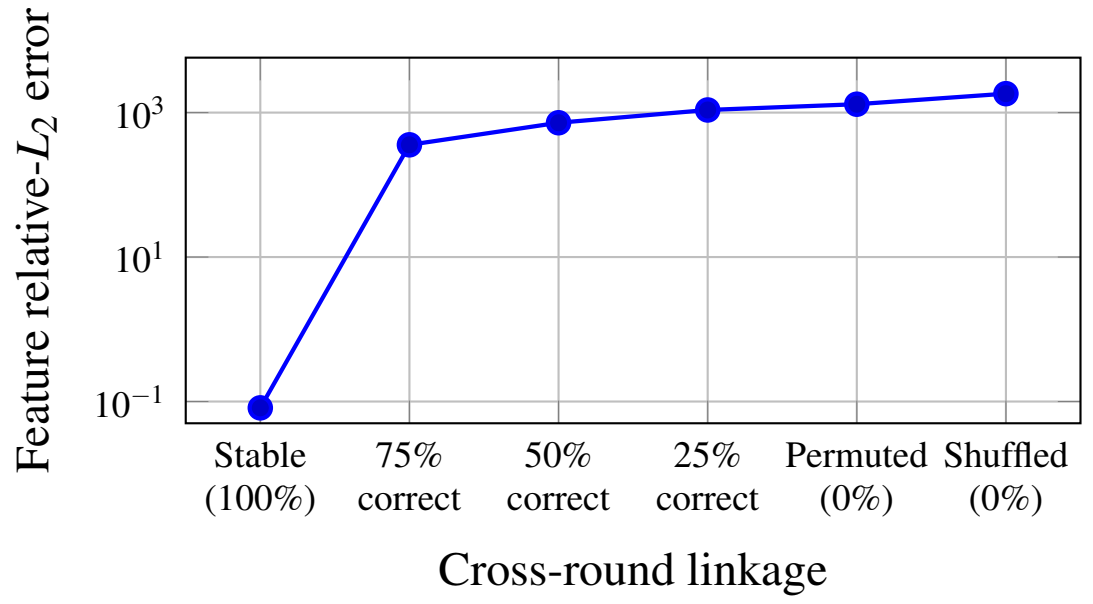

Holding $R_{\mathrm{rec}}=5$ fixed, Figure~\ref{fig:study1-linkage} isolates the effect of cross-round correspondence using the same reconstruction metric. Stable linkage yields relative-$L_2$ error $0.0817$. Reducing correct correspondence to 75\%, 50\%, and 25\% increases the error to 358.0, 723.5, and 1086.9, respectively; a fixed incorrect permutation yields 1306.1, and independent round-wise shuffling yields $1828.4\pm169.3$. Thus cross-round linkage strongly affects multi-round feature reconstruction but is not required for the direct first-round target disclosure above. In the \textsc{I-ER} control, the metadata-only activity probe remains at 0.0833.

\paragraph{Owner-side cost.}
Table~\ref{tab:study1-costs} shows that the HE burden is concentrated at the delegated owner. For $O_1$, peak RSS rises from 7.56\,MiB in \textsc{I-L} to approximately 87.6\,MiB in \textsc{I-OR}/\textsc{I-ER}, and transmitted volume from 0.009\,MiB to 16.26\,MiB. In contrast, $O_2$ and $O_3$ remain near 0.009\,MiB transmitted and approximately 11\,MiB peak RSS in the HE configurations. Configuration-level median end-to-end round latency is 0.402\,s for \textsc{I-L}, 0.441\,s for \textsc{I-P}, 4.931\,s for \textsc{I-OR}, and 3.185\,s for \textsc{I-ER}. These values characterize the tested placement and workload, not HE feasibility in general.

\begin{table}[t]
\centering
\setlength{\tabcolsep}{2.9pt}
\renewcommand{\arraystretch}{1.02}
\caption{Study-I owner resource profiles. Values are medians of three fresh-process repetitions.}
\label{tab:study1-costs}
 \scalebox{.83}{
\begin{tabular}{@{}llrrrr@{}}
\toprule
Owner & Config.  & CPU (ms) & RSS (MiB) & Sent/recv. (MiB) & Setup (s) \\
\midrule
$O_1$
& \textsc{I-L}  & 8.21  & 7.56  & 0.009/0.009 & $<$0.001 \\
& \textsc{I-P}  & 3.98  & 7.60  & 0.021/0.010 & 0.037 \\
& \textsc{I-OR} & 40.43 & 87.60 & 16.26/1.51  & 7.34 \\
& \textsc{I-ER} & 27.84 & 87.74 & 16.26/1.51  & 7.63 \\
\cmidrule(r){1-6}
$O_2$
& \textsc{I-L}  & 8.94 & 7.61  & 0.009/0.009 & $<$0.001 \\
& \textsc{I-P}  & 4.94  & 7.54  & 0.009/0.009 & $<$0.001 \\
& \textsc{I-OR} & 37.99  & 11.02 & 0.009/0.009 & $<$0.001 \\
& \textsc{I-ER} & 23.00  & 11.02 & 0.009/0.009 & $<$0.001 \\
\cmidrule(r){1-6}
$O_3$
& \textsc{I-L}  & 1.80  & 8.31  & 0.009/0.009 & $<$0.001 \\
& \textsc{I-P}  & 1.80  & 8.36  & 0.009/0.009 & $<$0.001 \\
& \textsc{I-OR} & 7.99  & 11.31 & 0.009/0.009 & $<$0.001 \\
& \textsc{I-ER} & 5.95  & 11.41 & 0.009/0.009 & $<$0.001 \\
\bottomrule
\end{tabular}
}
\end{table}

\subsection{Study II: Local Labels and Authenticated Messages}
\label{sec:type2-matched-study}

Study~II tests whether retaining labels at the owner prevents inference from the split boundary and whether authenticating executor messages establishes prescribed execution. Each record contains 36 motion features with 12 activity labels. The fixed partitions contain 504 learning, 144 utility-test, 72 attacker-auxiliary, and 108 attacker-target records.

We train a $36\!\rightarrow\!128\!\rightarrow\!64\!\rightarrow\!32\!\rightarrow\!12$ MLP for 10 rounds using ReLU, cross-entropy, SGD ($0.02$), minibatches of 32, one local epoch, and sample-weighted FedAvg. \textsc{II-L} is local FL and \textsc{II-S} a conventional split reference. In the U-shaped configurations, labels and loss remain at $O_1$. \textsc{II-U-E} keeps the first and last layers at the owner, \textsc{II-U-D} additionally moves the $128\!\rightarrow\!64$ layer to the owner, and \textsc{II-U-H} adds HMAC-SHA-256 authentication to the early-cut response.

\begin{figure}[t]
\centering
\begin{tikzpicture}
\begin{axis}[
    width=55mm,
    height=40mm,
    xlabel={Round},
    ylabel={Test balanced accuracy},
    xmin=1,
    xmax=10,
    xtick={1,2,3,4,5,6,7,8,9,10},
    ymin=0.08,
    ymax=0.48,
    tick label style={font=\scriptsize},
    label style={font=\small},
    grid=major,
]
\addplot+[
    mark=*,
    thick,
    error bars/.cd,
    y dir=both,
    y explicit,
]
coordinates {
    (1,0.1574)  +- (0,0.0561)
    (2,0.2222)  +- (0,0.0593)
    (3,0.2569)  +- (0,0.0783)
    (4,0.2824)  +- (0,0.0599)
    (5,0.3194)  +- (0,0.1028)
    (6,0.2894)  +- (0,0.0356)
    (7,0.3310)  +- (0,0.1329)
    (8,0.3519)  +- (0,0.0566)
    (9,0.3704)  +- (0,0.0607)
    (10,0.4051) +- (0,0.0328)
};
\end{axis}
\end{tikzpicture}
\caption{Study-II test balanced accuracy. All five configurations share the
same trajectory within numerical precision. Mean$\pm$SD across three seeds.}
\label{fig:study2-utility}
\end{figure}
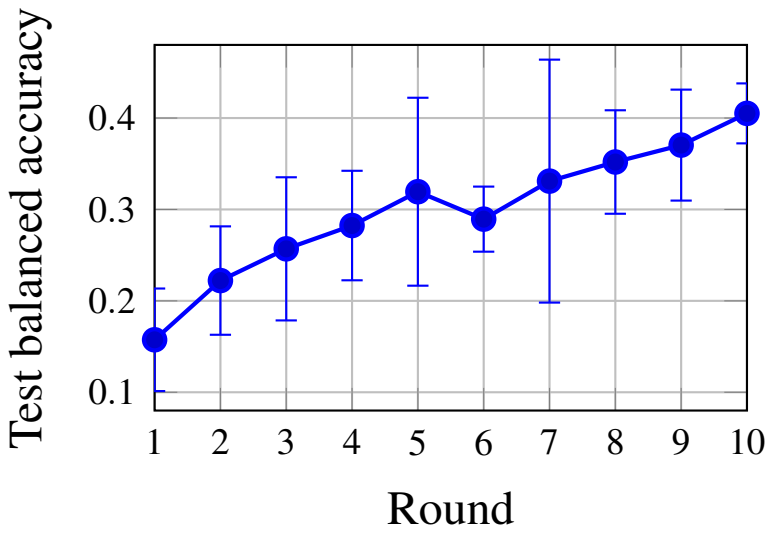

\paragraph{Matched utility and exposed boundary.}
All five configurations follow the same learning trajectory within numerical precision; at round~10, test balanced accuracy is $0.405\pm0.033$ and macro-F1 is $0.388\pm0.034$ (Figure~\ref{fig:study2-utility}). For the four split configurations, we probe executor-visible $z$, returned gradient $g$, and their concatenation $(z,g)$ using temporally disjoint auxiliary and target partitions. Each result is a separately trained finite-sample linear probe and therefore an attack-specific lower bound, not an upper bound on information in that view; in particular, the $(z,g)$ probe may underperform $g$ alone.

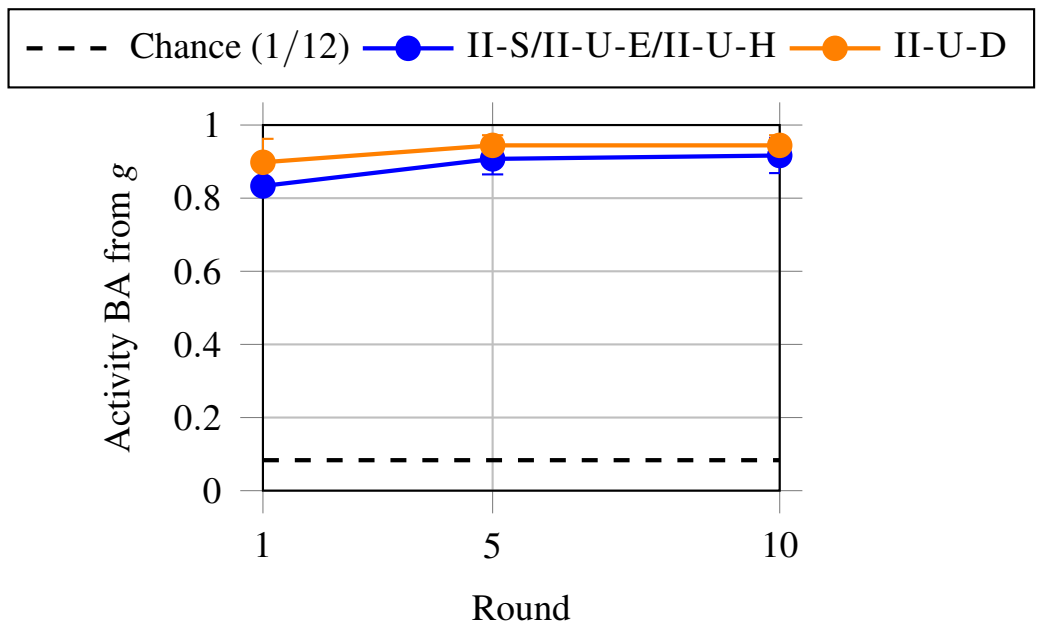
\begin{figure}[t]
\centering
\begin{tikzpicture}
\begin{axis}[
  width=5cm, height=4cm,
  grid=both,
  tick align=outside,
  scaled y ticks=false,
  tick label style={font=\scriptsize},
  label style={font=\scriptsize},
  yticklabel style={/pgf/number format/fixed},
  legend style={font=\scriptsize, at={(0.5,1.1)},anchor=south,legend columns = 5},
  xlabel={Round}, ylabel={Activity BA from $g$},
  xtick={1,5,10}, xmin=1, xmax=10,
  ymin=0,
     ymax=1
]

\addplot[
    black,
    dashed,
    mark=none,
    thick,
]
coordinates {
    (1,0.08333)
    (10,0.08333)
};
\addlegendentry{Chance ($1/12$)}

\addplot[
    blue,
    mark=*,
    thick,
    error bars/.cd,
    y dir=both,
    y explicit,
]
coordinates {
    (1,0.8333)  +- (0,0.0000)
    (5,0.9074)  +- (0,0.0424)
    (10,0.9167) +- (0,0.0481)
};
\addlegendentry{\textsc{II-S}/\textsc{II-U-E}/\textsc{II-U-H}}

\addplot[
    orange,
    mark=*,
    thick,
    error bars/.cd,
    y dir=both,
    y explicit,
]
coordinates {
    (1,0.8981)  +- (0,0.0642)
    (5,0.9444)  +- (0,0.0278)
    (10,0.9444) +- (0,0.0278)
};
\addlegendentry{\textsc{II-U-D}}

\end{axis}
\end{tikzpicture}
\caption{Activity inference from returned gradient $g$. The dashed line marks
$1/12$ chance; \textsc{II-S}, \textsc{II-U-E}, and \textsc{II-U-H}
overlap. Mean$\pm$SD across three seeds.}
\label{fig:study2-activity-leakage}
\end{figure}

At round~10, activity BA from $g$ is $0.917\pm0.048$ for \textsc{II-S}/\textsc{II-U-E}/\textsc{II-U-H} and $0.944\pm0.028$ for \textsc{II-U-D}, against $1/12=0.083$ chance. Activity BA from $(z,g)$ is $0.815\pm0.058$ and $0.843\pm0.032$, and subject BA is $0.523\pm0.064$ and $0.478\pm0.042$, respectively. Figure~\ref{fig:study2-activity-leakage} shows that activity leakage from $g$ is already high at round~1 and persists through round~10. For \textsc{II-U-D}, combining $g$ from rounds $\{1,5,10\}$ yields activity BA $0.991\pm0.016$.

\begin{table}[t]
\centering
\setlength{\tabcolsep}{.6pt}
\renewcommand{\arraystretch}{1.1}
\caption{Study-II round-10 privacy and owner-system results. Probe values are mean$\pm$SD over three seeds; system values are $O_1$ medians over three repetitions.}
\label{tab:study2-results}
\scalebox{.7}{
\begin{tabular}{@{}lrrrrrr@{}}
\toprule
& BA ($g$)
& BA ($z,g$)
& Subj. BA
& Sent/recv. (MiB)
& E2E (s)
& State (B) \\
\midrule

\textsc{II-L}
& -- & -- & --
& 0.60/0.60
& 0.868
& 26
\\

\textsc{II-S}
& $0.917{\pm}0.048$
& $0.815{\pm}0.058$
& $0.523{\pm}0.064$
& 1.99/1.87
& 1.277
& 26
\\

\textsc{II-U-E}
& $0.917{\pm}0.048$
& $0.815{\pm}0.058$
& $0.523{\pm}0.064$
& 2.19/2.08
& 1.279
& 26
\\

\textsc{II-U-D}
& $0.944{\pm}0.028$
& $0.843{\pm}0.032$
& $0.478{\pm}0.042$
& 1.46/1.36
& 1.221
& 26
\\

\textsc{II-U-H}
& $0.917{\pm}0.048$
& $0.815{\pm}0.058$
& $0.523{\pm}0.064$
& 2.19/2.09
& 1.336
& 968
\\

\bottomrule
\end{tabular}
}
\end{table}

\paragraph{Boundary placement and authentication.}
Table~\ref{tab:study2-results} shows different communication, latency, and state costs across the split configurations. Relative to \textsc{II-S}, \textsc{II-U-E} increases communication from 1.99/1.87\,MiB to 2.19/2.08\,MiB with nearly unchanged latency (1.277\,s versus 1.279\,s), while the tested probe results are identical. Moving from \textsc{II-U-E} to \textsc{II-U-D} reduces transmitted and received traffic by approximately 33\% and 35\%; the privacy-probe point estimates differ only modestly relative to the observed seed variation. Median latency changes from 1.279\,s to 1.221\,s, so neither cut dominates across the measured dimensions.

\textsc{II-U-H} authenticates the executor response with task, round, batch, model-state, sequence, nonce, and payload context. All 24 payload-modification, replay, and context-mismatch controls are rejected, but all 180 tested deviations by the legitimate key holder (partial execution, stale reuse, gradient negation, and substitution) are accepted with fresh valid authentication. Stale reuse reduces executor active computation by a median 56.4\% while preserving a valid HMAC. Thus, under the stated key-distribution assumption, HMAC establishes message integrity and authenticates possession of the shared key. It binds the message to the supplied context but does not establish that the executor performed the prescribed computation.

In the system measurements, authentication raises owner active CPU from 0.313\,s to 0.318\,s and end-to-end latency from 1.279\,s to 1.336\,s; persistent state rises from 26\,B to 968\,B, with no measured change in peak RSS.

\paragraph{Feasibility under memory budgets.}
A supplementary local-training analysis scales the Study-II MLP under explicit memory budgets. The $16\times$ model (approximately 2.7 million parameters) fails to train under 16 and 64\,MiB budgets but fits within 256\,MiB. The $32\times$ model (approximately 10.6 million parameters) fails under a 256\,MiB budget. These results illustrate how workload size can make local training infeasible even when the smaller model fits the device. However, the delegated resource measurements in Tables~\ref{tab:study1-costs} and~\ref{tab:study2-results} concern smaller workloads. They do not establish that protected delegation of the scaled models fits the same memory budgets.

\paragraph{Takeaway.}
The studies show how protection of an individual artifact can leave other parts of the executor view exposed. Cross-round linkage affects feature reconstruction in Study~I, whereas direct first-round target disclosure under zero initialization does not require linkage. Study~II shows that privacy and attribution gaps can occur within the same delegated configuration. In \textsc{II-U-H}, the gradient reveals activity labels with a balanced accuracy of $0.917\pm0.048$, while all 180 tested key-holder deviations pass authentication. At the same time, owner transmissions increase from 0.60\,MiB under local FL to 2.19\,MiB under authenticated delegation. Authentication therefore adds owner-side costs without establishing prescribed execution or preventing the tested inference. The studies illustrate these interactions under matched configurations but do not establish that delegation enables otherwise infeasible participation. These conclusions are bounded to the tested workloads and threat models.

\section{Research Directions, Limitations, and Conclusion}
\label{sec:agenda}

\subsection{Research Directions}

The five cross-axis findings suggest concrete requirements for designing and evaluating Delegated FL. The common objective is to establish privacy, attribution, and owner feasibility over the same owner-attributed delegated transition rather than evaluate each property under a different boundary or configuration.

\noindent\textbf{Accumulated-view privacy.}
Privacy evaluation should specify the protected secret and the information available to the observer or coalition over the execution horizon. This includes visible artifacts, auxiliary information, and permitted leakage. Repeated and linkable observations should be included when available. For adaptive delegation, the evaluation should also account for changes in the observable view as the execution boundary moves across rounds. Formal guarantees should remain scoped to their stated adversary and leakage function, while empirical evaluations should identify the tested attacks and the portions of the accumulated view that remain unassessed.

\noindent\textbf{Delegated-attribution coverage.}
Assurance claims should identify which predicates in Section~\ref{sec:delegated-attribution} are established and how their verification affects admission. Proofs, sampled audits, TEEs, behavioral checks, and incentives establish different relations under different assumptions and should not be treated as interchangeable. Payment contingent on verified service~\cite{abadi2023recurring}, for example, addresses rational work evasion only under an explicit utility model. For deterministic training, replicated execution offers an alternative in which the owner accepts a result only when independently assigned executors return matching outputs. Assuming at least one honest executor and identical authorized inputs and execution conditions, this can provide lightweight result verification at the owner. However, it introduces additional communication and duplicated computation, and may expose owner information to more parties. For stochastic training, the authorized policy should define permitted randomness and approximation so that legitimate execution can be distinguished from unauthorized deviations.

\noindent\textbf{Complete owner-resource accounting.}
Feasibility claims for constrained owners should identify the target device and network class and measure the complete residual workflow using the resource dimensions in Section~\ref{sec:complexity}. The evaluation should account for the cost of protection and verification alongside any retained learning. Furthermore, setup and recovery costs should be included rather than excluded from the owner budget. To establish that delegation enables otherwise infeasible participation, local and delegated execution should be evaluated against the same owner budget under matched workloads and security requirements. For adaptive systems, both cumulative resource consumption and peak demands should be reported, since reductions in average cost do not necessarily establish feasibility.

\noindent\textbf{Explicit adversary and administrative boundaries.}
Future designs should state which roles may be malicious or colluding and identify the administrative boundaries under which their guarantees hold. Executor--aggregator collusion may enlarge the observable view, while verifier compromise may invalidate admission evidence. Malicious owners introduce a distinct input-integrity problem. An executor may correctly perform the authorized learning procedure on poisoned owner data. Satisfying the input-binding and prescribed-work predicates does not establish that the authorized data are benign. Furthermore, permission to delegate computation should be distinguished from permission to disclose or process the underlying data. This distinction is particularly relevant when the executor operates under a different administrative or regulatory authority. A design need not address every adversarial relation, but its supported and excluded cases should be explicit.

\noindent\textbf{Aligned benchmark bundles.}
Comparative evaluations should use matched workloads and measurement boundaries across representative constrained-device and network classes. Each run should associate its transcript and resource measurements with the corresponding security configuration and threat model. Baseline and protected variants should evaluate model utility alongside privacy and attribution under the same execution conditions. The corresponding owner costs should also be measured. Such bundles would enable systematic comparison without collapsing the three dimensions into a single security or efficiency score.

\subsection{Limitations}
\label{sec:limitations}

The conclusions of this SoK are bounded by the reviewed corpus, the evidence available in the primary sources, and the controlled conditions of the two studies. The corpus is defined by a semantic inclusion rule applied across heterogeneous terminology spanning federated, split, edge-assisted, and outsourced learning. Our search emphasizes delegated execution with security or privacy considerations. Although we supplemented the database searches with backward and forward snowballing, relevant systems may have been missed. The corpus also includes both explicitly IoT-oriented systems and more general designs that instantiate the same owner--executor separation. The resulting synthesis therefore characterizes mechanisms and evidence relevant to Delegated FL rather than their prevalence in deployed IoT systems. Missing evidence is treated as unreported rather than as evidence that the corresponding property is absent.

The controlled studies instantiate recurring corpus patterns but do not reproduce individual audited systems or constitute a representative benchmark of Delegated FL. Both use PAMAP2-derived tasks with small models and fixed owner assignments on one five-machine deployment. Study~I's near-exact first-round target recovery follows specifically from zero initialization and disclosure of the opened residual. It does not imply that CKKS-based delegation generally reveals training targets. Multi-round feature reconstruction depends on the model trajectory, cross-round linkage, and numerical conditioning of the accumulated equations. Furthermore, low aggregate reconstruction error does not necessarily establish accurate recovery of individual features. Study~II evaluates activity and subject inference using specified linear probes and temporally disjoint auxiliary data. The reported accuracies do not provide an upper bound on leakage under other attacks. The numerical results therefore apply to the tested configurations and threat models.

The resource measurements characterize the tested Raspberry Pi owners and the corresponding executor, aggregator, and network deployment. The main experimental workloads remain locally trainable on these devices. Consequently, the measured costs illustrate how delegation redistributes owner burden rather than establish that protected delegation enables otherwise infeasible participation. Different device budgets and workloads may change the relative feasibility of local and delegated execution. Direct energy is not reported because reproducible power instrumentation was unavailable. Furthermore, the studies focus on the owner--executor boundary and do not evaluate executor--aggregator collusion, malicious owners, verifier compromise, denial-of-service, or other multi-party failures. We also do not empirically compare alternative execution-assurance mechanisms such as zero-knowledge proofs or TEEs. The studies therefore provide controlled evidence for the tested corpus-grounded patterns rather than an end-to-end security or feasibility evaluation of Delegated FL deployments.

\subsection{Conclusion}
\label{sec:conclusion}

Delegated FL separates ownership of a federated contribution from execution of the learning that produces it. Our systematization shows that this separation creates three coupled questions: what the executor can infer from its accumulated view, which relations bind delegated execution to the owner's authorized contribution, and whether the complete residual workflow remains feasible for the owner.

Across the reviewed systems, privacy and attribution mechanisms frequently establish guarantees over different parts of the delegated transition. The deployment evidence does not necessarily characterize the owner-side cost of realizing these guarantees. The resulting composition gap does not imply that the individual mechanisms fail to provide their stated guarantees. Rather, these guarantees and their supporting evidence do not necessarily establish the same end-to-end claim. The controlled studies instantiate several recurring corpus patterns and illustrate this distinction under matched experimental conditions. By using the owner-attributed delegated transition as the unit of analysis, our framework identifies where assurance remains incomplete and where further evidence is required to establish private, attributable, and feasible delegation.

\appendix

\section{Ethical Considerations}
\label{app:ethics}

We considered this work using a stakeholder-based analysis informed by beneficence, respect for persons, justice, and respect for law and public interest. Relevant stakeholders include PAMAP2 participants and data stewards, authors of audited systems, IoT owners and end users, FL operators and vendors, security researchers and potential attackers, and the research team.

The controlled experiments use the publicly available PAMAP2 dataset and researcher-controlled computing and network infrastructure. They do not interact with production systems, external users, or third-party services. For dataset participants, the main risk is demonstrating additional inference capability from released data. We therefore use only the published dataset, perform no external identity linkage, and restrict conclusions to the stated experimental conditions.

For authors of audited systems, an ethical risk is presenting an unreported or out-of-scope property as a system failure. We mitigate this through conservative \emph{Not Reported} coding, source locators, and claims bounded to the guarantees and evidence provided by the primary sources. Absence of evidence is not treated as evidence that a system violates a property it did not claim.

For IoT owners, end users, operators, and vendors, the principal risk is false confidence if protection of one artifact or relation is interpreted as end-to-end assurance. We therefore make observer, trust, configuration, and exclusion assumptions explicit and limit findings to the execution paths actually evaluated.

The reconstruction and work-evasion analyses also have dual-use potential: they can help identify assurance gaps but could be adapted offensively. We reduce this risk by evaluating them only in controlled environments and by releasing no production credentials, third-party targets, or tooling for accessing external systems. This does not eliminate the possibility that the techniques could be reused after publication.

We judged the research appropriate to conduct because it uses previously released data and researcher-controlled systems without interacting with unsuspecting users or exploiting production systems. We also judged publication appropriate because exposing these assurance gaps can help designers and evaluators identify and mitigate them. The remaining dual-use risk is limited through bounded claims and a reproducibility artifact scoped to the reported experiments and distributed subject to applicable dataset, software, and artifact licensing terms.

\section{Open Science}
\label{app:open-science}

The artifacts supporting the two controlled studies are available at: 
%
\url{https://github.com/AlghamdiHamed/SoK_DelegatedFL}

The artifact contains the implementations of Studies~I and~II, experiment configurations, execution scripts, attack and work-evasion code, dependency information, configuration manifests, and the processed results used to produce the reported measurements and tables.

For Study~I, the artifact includes the Go/Lattigo implementation, CKKS configuration, delegated-execution variants, correctness checks, accumulated-view reconstruction analysis, and resource-measurement scripts. For Study~II, it includes the PyTorch implementation, split-boundary configurations, privacy probes, HMAC controls, prescribed-work deviation tests, and resource-measurement scripts.

The repository also contains instructions for reproducing the experiments and identifying the configurations corresponding to the results reported in Section~\ref{sec:costs}. The PAMAP2 dataset is not redistributed; the artifact provides preprocessing instructions and the configuration required to construct the experimental partitions from the publicly available dataset.

\section*{Acknowledgments}
Hamed Mohammed Alkharsh gratefully acknowledges Al-Baha University, Saudi Arabia, for supporting his doctoral studies through a scholarship. Aydin Abadi was supported in part by Kuwait Foundation for the Advancement of Sciences (KFAS) under project code: PA24-6TE-2493.

\bibliographystyle{plainurl}
\bibliography{ref}

\cleardoublepage


\newcommand{\AuditSupported}{\textsc{Supported}}
\renewcommand{\AuditLimited}{\textsc{Limited}}
\newcommand{\AuditUnsupported}{\textsc{Unsupported}}

\renewcommand{\AuditNR}{\textsc{Not reported}}

\clearpage
\onecolumn
\raggedbottom
\section{Per-System Property and Evidence Audit}
\label{app:corpus}
\label{app:multiaxis_audit}

\paragraph{Coding interpretation.}
Tables~\ref{tab:audit_structure_trust}--\ref{tab:audit_client_burden} provide the per-system record used in Sections~\ref{sec:model}--\ref{sec:findings}. For delegate-view privacy, delegated attribution, and constrained-owner feasibility, the audit separates \emph{design support} from \emph{evidence status}. Design support records whether the audited workflow establishes the property under its stated assumptions: \AuditSupported{}, \AuditLimited{}, \AuditUnsupported{}, or \AuditNA{}. When the source is insufficient to determine support, the design-support entry is \AuditNR{}. Evidence status records how the coded judgment is substantiated: formal security proof, analytical argument, empirical privacy-attack evaluation, prototype measurement, real-device evaluation, simulation only, author claim only, or \AuditNR{}. These fields are independent. In particular, \AuditNR{} never means that a design is inefficient or insecure; it means that the paper does not supply enough information to decide. Conversely, an \AuditUnsupported{} property remains unsupported even if the system has extensive performance measurements.

Administrative ownership is recorded only when the paper makes it clear. Otherwise, the entry remains \AuditNR{}, even when the physical roles appear distinct. The entries are survey judgments based on the reviewed papers, not terminology necessarily used by their authors.

For direct reconciliation with the main corpus table, every partial-delegation entry states the underlying codes in the same grouped order: delegate adversary and executor--aggregator collusion (DA, C); label protection, cryptographic guarantee, differential privacy, inference-attack evidence, and cross-round analysis (L, CG, DP, IA, XR); freshness, prescribed-work assurance, and consequential admission (Fr, W, AD); and retained learning work, returned-boundary combination, additional owner work, real constrained-device evidence, and joint peak-memory/communication evidence (LW, BX, AW, IoT, M+C). Here \AuditSupported{}, \AuditLimited{}, \AuditUnsupported{}, \AuditNR{}, and \AuditNA{} correspond to S, L, N, NR, and NA, respectively. Physical PDF pages are used for partial-delegation source locators.

For serverless systems, full-delegation classification applies only to the explicitly evaluated public-cloud FaaS branch in which the data-owner client function is executed by the named cloud platform. Self-hosted or endpoint-local branches are not treated as delegated merely because they use a FaaS abstraction.

\scriptsize
\setlength{\tabcolsep}{2.2pt}
\renewcommand{\arraystretch}{1.05}

\begin{longtable}{
  >{\raggedright\arraybackslash}p{0.07\linewidth}
  >{\raggedright\arraybackslash}p{0.135\linewidth}
  >{\raggedright\arraybackslash}p{0.305\linewidth}
  >{\raggedright\arraybackslash}p{0.195\linewidth}
  >{\raggedright\arraybackslash}p{0.235\linewidth}
}
\caption{Transition structure, roles, and trust boundary.}
\label{tab:audit_structure_trust}\\
\toprule
\textbf{Ref.} & \textbf{Delegation / primary structure} & \textbf{Data owner; delegate; aggregator} & \textbf{Verifier / authority} & \textbf{Collusion and administrative ownership} \\
\midrule
\endfirsthead
\toprule
\textbf{Ref.} & \textbf{Delegation / primary structure} & \textbf{Data owner; delegate; aggregator} & \textbf{Verifier / authority} & \textbf{Collusion and administrative ownership} \\
\midrule
\endhead
\bottomrule
\endfoot
\cite{dong2020offloading} & Full / full training & Client; TEE-enabled edge server; FL server & TEE runtime; attestation details \AuditNR{} & Delegate--aggregator collusion \AuditNR{}; TEE operator and administrative ownership \AuditNR{} \\
\cite{jijin2022smart} & Full / data or task offloading & Client; seed and v-FAP service nodes; FL coordinator & Smart contracts and blockchain log & Collusion \AuditNR{}; operator ownership \AuditNR{} \\
\cite{zhang2025zkfhed} & Full / full training & Client; FHE/ZK workers; aggregator & Request, training, and decryption proof verifiers; key holders & Collusion threshold follows the cryptographic setup; deployment ownership \AuditNR{} \\
\cite{han2025outsourced} & Full / full training & User/data owner; private-cloud trainer; FL server & Data owner using sensitive-sample checks & Collusion and administrative control of the private cloud \AuditNR{} \\
\cite{lin2022td3} & Full / data or task offloading & Device; selected edge/fog node; FL server & None reported & Collusion and administrative ownership \AuditNR{} \\
\cite{marx2023ww} & Full / full training & Client/data owner; Layer-II MPC cluster servers; Layer-I global MPC servers & MPC protocols and ideal functionalities; no separate external verifier & Clients trust each cluster not to collude entirely; the mixed centralized-adversary model requires at least one honest entity per set; prototype CrypTen uses two semi-honest servers and a trusted helper; administrative ownership \AuditNR{} \\
\cite{chadha2020towards} & Full / full training & Data-owner client; commercial FaaS client function in the GCF branch; FL server & Weights-Updater checks invocation completion/errors and retries failures; no computation verifier & Commercial GCF and self-hosted/on-device FaaS branches coexist; contractual administrative independence, host trust, and collusion \AuditNR{} \\
\cite{li2021quantum} & Full / delegated gradient computation & Client/data owner; remote quantum server; shared-model updater or aggregator not separately specified & Client drives UBQC and derives the returned result; no separate computation or admission verifier & Quantum server is explicitly untrusted; client--server administrative ownership and any executor--aggregator collusion relationship \AuditNR{} \\
\cite{grafberger2021fedless} & Full / full training & Client administrator/data holder; commercial FaaS client function; FedLess FL server/aggregator & AWS Cognito/JWT authenticates invocation; controller monitors completion; no computation verifier & Data holder deploys the function; commercial and self-hosted FaaS coexist; host trust, collusion, and administrative independence \AuditNR{} \\
\cite{elzohairy2022fedlesscan} & Full / full training & Data-holder client; GCF client function; FedLess controller and OpenFaaS aggregator & Aggregator checks update age against the authorized round and discards over-age updates; no prescribed-work verifier & GCF clients with a self-hosted aggregator; malicious clients are explicitly out of scope; host trust, collusion, and administrative independence \AuditNR{} \\
\cite{xu2025sgboost} & Partial / vertical tree-training subcomputation & Passive VFL data owner; CSP and active party execute protected subroutines; active party coordinates training & Key/share-holding protocol parties; no malicious-computation verifier & DA=\AuditSupported{}; C=\AuditUnsupported{}. CSP and participants are honest-but-curious and explicitly non-colluding; administrative ownership \AuditNR{} (PDF pp.~3--4, Sec.~III). \\
\cite{thapa2022splitfed} & Partial / model-partitioned & Client; main split-learning server; federated server & None reported & DA=\AuditSupported{}; C=\AuditNR{}. The main server is treated as curious in the privacy variants; its administrative and collusion relationship with the federated server is not specified (PDF pp.~2--4, Fig.~1 and Algorithms~1--2). \\
\cite{wu2023federated} & Partial / vertically split data/label training & Vehicle and RSI respectively hold data and labels and execute complementary submodels; FL coordinator aggregates & None reported & DA=\AuditSupported{}; C=\AuditUnsupported{}. The complementary party is an explicit privacy attacker, while the analyzed collusion case is unsupported by the stated model (PDF pp.~3--6, Secs.~III--IV and Algorithms~1--5). \\
\cite{quan2023hiersfl} & Partial / hierarchical model partition & Client; mobile edge server (MES); cloud aggregator & None reported & DA=\AuditNR{}; C=\AuditNR{}. Neither an adversarial MES nor MES--cloud collusion/administrative independence is specified (PDF pp.~3--4, Fig.~1 and Algorithm~1). \\
\cite{stephanie2023dtasfl} & Partial / asynchronous model partition & Hospital/IoMT owner and edge component; cloud-side split executor and asynchronous aggregator & Secret-sharing participants protect aggregation values; no training verifier & DA=\AuditSupported{}; C=\AuditNA{}. The cloud is the delegated training and aggregation role in the audited path, so there is no distinct executor--aggregator pair; ownership otherwise \AuditNR{} (PDF pp.~4--8, Sec.~IV). \\
\cite{meng2025pptadi} & Partial / U-shaped protected model partition & AIoT owner device; main server; aggregation TTP & TTP for protected aggregation; no prescribed-work verifier & DA=\AuditSupported{}; C=\AuditNR{}. The main server is semi-honest; the TTP/non-collusion and administrative relationships are only partially specified (PDF pp.~5--6, Algorithms~1--3 and Sec.~4.3). \\
\cite{stephanie2025weight} & Partial / asynchronous partial model partition & Hospital/IoMT owner and edge component; cloud-side split executor/aggregator & Sharing participants protect selected aggregation messages; no training verifier & DA=\AuditNR{}; C=\AuditNA{}. No adversarial cloud executor is specified, and the same cloud workflow performs delegated training and aggregation (PDF pp.~7--11, Algorithms~1--3). \\
\cite{zhao2023usfl} & Partial / U-shaped model partition & Client; split server that also coordinates server-side/federated processing & None reported & DA=\AuditSupported{}; C=\AuditNA{}. The server is explicitly potentially malicious, but no separate aggregator exists in the audited workflow (PDF pp.~2--3, Sec.~II-A and Algorithm~1). \\
\cite{chen2025llmiot} & Partial / model-partitioned LoRA fine-tuning & IoT device; edge server that performs the upper-part training and aggregation & None reported & DA=\AuditLimited{}; C=\AuditNA{}. The edge observer is used for reconstruction-leakage analysis but is not given a full malicious/work-evasive model; no separate aggregator exists (PDF pp.~2--4, system model and privacy metric). \\
\cite{huang2025dpmora} & Partial / adaptive model partition & End device; edge server that performs the server-side submodel and federated coordination & Leakage estimator only; no computation verifier & DA=\AuditSupported{}; C=\AuditNA{}. The edge is explicitly considered in reconstruction attacks and is also the aggregation role (PDF pp.~3--5, system/privacy model). \\
\cite{chen2024shieldtse} & Partial / model-partitioned traffic estimation & RSU/data owner; cloud executor and cloud coordinator/aggregator & None reported & DA=\AuditSupported{}; C=\AuditNA{}. The honest-but-curious cloud is both delegated executor and aggregator (PDF p.~5, Sec.~III-C; Fig.~1). \\
\cite{hou2025verifyvfl} & Partial / outsourced VFL gradient subroutine & VFL data owners; computing party; no distinct FL aggregator in the verified subroutine & Data owners verify with public parameters supplied by a trusted authority & DA=\AuditSupported{}; C=\AuditNA{}. The computing party and colluding passive parties are adversarially modeled, but executor--aggregator collusion is structurally inapplicable; TA and administrative ownership remain assumptions (PDF pp.~4--5, Sec.~IV-C). \\
\cite{latif2025hmac} & Partial / model-partitioned & CIoT client; split-learning server; federated server & Message endpoints verify HMACs under a shared-key setup & DA=\AuditNR{}; C=\AuditNR{}. Tampering attackers are modeled on exchanged messages, but the delegated executor itself and its relationship with the FL server are not specified as adversarial/colluding (PDF pp.~2--4, proposed framework and Algorithm~1). \\
\cite{okegbile2025novel} & Partial / split federated semantic learning & Physical-twin/edge-side owner path; cloud-side model executor and cloud aggregation/validation path & Token checks and blockchain validators; authority details \AuditNR{} & DA=\AuditNR{}; C=\AuditNA{}. The cloud-side path combines delegated learning and aggregation; no adversarial cloud-executor model is stated, and administrative ownership is \AuditNR{} (PDF pp.~4--7, Secs.~III--IV and Algorithm~1). \\
\cite{cox2022aergia} & Partial / peer-assisted layer offloading & Weak client; stronger peer client; federator schedules and aggregates & Federator signs schedules and discards late-round results; SGX protects selected profiling metadata only & DA=\AuditNR{}; C=\AuditNR{}. The stronger peer is not explicitly adversarial, and peer--federator collusion/ownership is not analyzed (PDF pp.~4--8, Secs.~3--4 and Algorithms~1--2). \\
\cite{xia2022hsfl} & Partial / hierarchical model partition & Client; edge split server; cloud aggregator & None reported & DA=\AuditNR{}; C=\AuditNR{}. The edge training assistant and edge--cloud relationship are not adversarially modeled (PDF pp.~2--4, Sec.~II and Algorithm~1). \\
\cite{wu2025hfsl} & Partial / heterogeneous model partition & Heterogeneous client; split server; FL server & None reported & DA=\AuditNR{}; C=\AuditNR{}. No adversarial split-server or executor--aggregator collusion model is stated (PDF pp.~5--19, Secs.~3--4 and Algorithms~1--3). \\
\cite{fei2025privacy} & Partial / data-partitioned local/RSU training (DRWB; author-assigned short name) & Vehicle/data owner; RSU trains the offloaded non-private data portion; cloud aggregates both locally and remotely trained parameters & None reported & DA=\AuditUnsupported{}; C=\AuditUnsupported{}. The performance-optimization model assumes cooperative RSU/cloud execution and provides no adversarial or collusion resistance (PDF pp.~4--10, Secs.~III--VI). \\
\cite{he2023computation} & Partial / data-partitioned local/MEC training & Mobile device/data owner; MEC server trains the offloaded data portion; aggregation server forms the global model & None reported & DA=\AuditUnsupported{}; C=\AuditUnsupported{}. MEC/aggregation entities are cooperative optimization components, with no adversarial or collusion protection (PDF pp.~4--11, Secs.~II--IV and DDPG algorithm). \\
\end{longtable}

\begin{longtable}{
  >{\raggedright\arraybackslash}p{0.05\linewidth}
  >{\raggedright\arraybackslash}p{0.17\linewidth}
  >{\raggedright\arraybackslash}p{0.18\linewidth}
  >{\raggedright\arraybackslash}p{0.21\linewidth}
  >{\raggedright\arraybackslash}p{0.15\linewidth}
  >{\raggedright\arraybackslash}p{0.18\linewidth}
}
\caption{Boundary artifacts and delegate-view privacy. Design support and evidence status are coded independently.}
\label{tab:audit_artifacts_privacy}\\
\toprule
\textbf{Ref.} & \textbf{Boundary artifacts} & \textbf{Privacy objective} & \textbf{Design support} & \textbf{Evidence status} & \textbf{Residual leakage} \\
\midrule
\endfirsthead
\toprule
\textbf{Ref.} & \textbf{Boundary artifacts} & \textbf{Privacy objective} & \textbf{Design support} & \textbf{Evidence status} & \textbf{Residual leakage} \\
\midrule
\endhead
\bottomrule
\endfoot
\cite{dong2020offloading} & Encrypted client data; plaintext inside TEE; local update; timing metadata & Hide client records and training state from the edge host outside the enclave & \AuditLimited{}: TEE isolation supports host-side confidentiality under the hardware boundary; it does not hide plaintext from the enclave and does not cover all side channels or metadata & Analytical argument; simulation only for offloading performance; attestation and side-channel evaluation \AuditNR{} & Updates, timing, and access metadata remain visible \\
\cite{jijin2022smart} & Raw data; initial parameters; local outcomes; transactions; logs & Hide owner data from service-node delegates & \AuditUnsupported{}: access control, blockchain logging, and orchestration do not conceal raw data from the authorized service nodes & Author claim only for confidentiality; prototype measurement for orchestration; delegate-leakage evaluation \AuditNR{} & Service nodes see raw data and training state; protocol logs expose metadata \\
\cite{zhang2025zkfhed} & Encrypted data and updates; ZK circuits; proofs; hashes; logs & Hide training data and update contents from workers and infrastructure & \AuditSupported{} for the encrypted artifacts under the stated FHE/ZK setup; \AuditLimited{} for metadata and any relation outside the proved circuit & Formal security proof; prototype measurement & Circuit shape, ciphertext/proof sizes, hashes, participation, timing, and setup metadata remain visible \\
\cite{han2025outsourced} & Training batches; privacy budgets; noisy gradients; updates; sensitive samples & Protect updates from the aggregator and owner data from the private-cloud delegate & \AuditLimited{}: perturbation supports update privacy against the aggregator; owner-to-delegate confidentiality is \AuditUnsupported{} because the trainer receives the batches & Analytical argument; prototype measurement; delegate-side privacy attack evaluation \AuditNR{} & The delegate sees training batches and related state; behavioral checks do not reduce that view \\
\cite{lin2022td3} & Task data; parameters; node-selection and resource metadata & Hide task data from the selected edge/fog node & \AuditUnsupported{}: no confidentiality mechanism is provided at the delegation boundary & Privacy evidence \AuditNR{}; simulation only for scheduling performance & The selected delegate sees task data and associated metadata \\
\cite{marx2023ww} & Secret shares of client records and labels; secret-shared current, cluster, and global models; client-selection, resharing, and aggregation messages & Hide client data and the global model from any entity or colluding set below the configured MPC corruption threshold & \AuditSupported{} under the stated MPC instantiation and corruption threshold; \AuditLimited{} once the threshold is exceeded and for access-pattern, participation, size, and timing metadata & Formal simulation-based security proof (Theorem~1); CrypTen/AWS prototype and poisoning-attack evaluation; delegate-side privacy-attack evaluation \AuditNR{} & Collusion beyond the configured threshold defeats confidentiality; cluster membership, selection, message sizes, timing, and authorized output/inference metadata remain visible \\
\cite{chadha2020towards} & Current weights; model/task configuration; client-specific training data and labels; updated weights; optimizer state; client ID and invocation status & Protect client data from other clients and the FL server; no stated confidentiality objective against the FaaS host & \AuditUnsupported{} for delegate-view record/label confidentiality: the FaaS client function loads plaintext training data; per-client bucket isolation prevents other-client access but does not conceal data from the executing platform & Prototype measurement across GCF, OpenWhisk, and OpenFaaS; delegate-leakage evaluation, DP guarantee, and cryptographic confidentiality proof \AuditNR{} & The FaaS executor processes records, labels, model state, and optimizer state; client identity, timing, and invocation metadata remain visible \\
\cite{li2021quantum} & Randomized qubits encoding client inputs and labels; adaptive measurement instructions and outcomes; clipped/noisy gradients; public shared-model parameters & Blind the delegated inputs, outputs, and computation from the quantum server and reduce leakage from released gradients & \AuditSupported{} for UBQC blindness against the untrusted quantum server under the stated client capability; \AuditLimited{} for released-gradient privacy because no formal privacy accounting is supplied and the paper acknowledges residual model/label leakage & Analytical inheritance from UBQC; numerical simulation and gradient-reconstruction attack evaluation; no real quantum-device experiment & The public model and noisy updates remain observable; training curves can leak label information, and client identity, interaction pattern, timing, and message sizes remain linkable \\
\cite{grafberger2021fedless} & JWTs and scoped credentials; current model and hyperparameters; local data and labels; individual update; optional DP noise; client-registry and timing metadata & Protect client records and update access while supporting commercial FaaS execution & \AuditLimited{}: TLS, scoped parameter-server access, and optional record-level LDP protect transport, cross-client access, and the specified release; executor-view data/label confidentiality and aggregator-view update confidentiality are \AuditUnsupported{} because the function trains on plaintext and secure aggregation is not used & Analytical LDP argument through the stated TensorFlow Privacy mechanism; cross-cloud prototype measurement; delegate-side leakage, side-channel, and cross-round linkability evaluation \AuditNR{} & The FaaS executor sees training records and labels; the aggregator can access individual updates; client identity, function timing, dataset location, and external privacy-budget state remain linkable \\
\cite{elzohairy2022fedlesscan} & Current model; local data and labels; round-tagged update; training time; missed-round and cooldown history; selection and staleness metadata & Protect client data while mitigating serverless stragglers & \AuditUnsupported{} at the delegation boundary: the client function loads plaintext data and labels; inherited FedLess DP is mentioned but no FedLesScan-specific executor-view protection is established or evaluated & GCF prototype measurement and simulated straggler evaluation; privacy proof, leakage attack, and cross-round linkability evaluation \AuditNR{} & The executor sees client records, labels, and model state; persistent per-client timing, missed-round, cooldown, selection, and update-age histories are linkable \\
\cite{xu2025sgboost} & Secret-shared bucket indicators; encrypted gradients/model values; random coins; ciphertexts & Protect labels and vertical tree-training values from the CSP and active party & L=\AuditSupported{}; CG=\AuditSupported{}; DP=\AuditUnsupported{}; IA=\AuditSupported{}; XR=\AuditLimited{}. Cryptography protects the specified values under honest-but-curious non-collusion; no DP is used and repeated-transcript protection is only protocol-relative & Formal/analytical cryptographic security analysis and empirical privacy/security evaluation (PDF pp.~4 and 9--13, threat model and Secs.~VI--VII) & Access patterns, sizes, repeated transcripts, participant identity, and authorized output/model leakage remain; collusion violates the stated guarantee \\
\cite{thapa2022splitfed} & Smashed data; labels in SFLV1/default flow; activation gradients; client/server model updates & Protect selected activations/updates while retaining split execution & L=\AuditUnsupported{}; CG=\AuditUnsupported{}; DP=\AuditSupported{}; IA=\AuditUnsupported{}; XR=\AuditSupported{}. The paper's DP mechanisms formally govern specified repeated releases, but the audited default architecture does not keep labels from the main server & Formal DP construction/composition and prototype performance evaluation (PDF pp.~2--7, Algorithms~1--2 and privacy variants); no paper-run reconstruction/inference attack & Default-flow labels, unprotected intermediates, model state, identities, sizes, and timing remain visible outside the DP target \\
\cite{wu2023federated} & Smashed/intermediate data; cut-layer and local gradients; sample identifiers linking vehicle data to RSI labels & Protect data and labels against complementary-party reconstruction/inference & L=\AuditUnsupported{}; CG=\AuditUnsupported{}; DP=\AuditSupported{}; IA=\AuditSupported{}; XR=\AuditSupported{}. Raw labels are not local to the data owner; the paper instead protects specified intermediate releases through composed DP and evaluates named attacks & Formal DP analysis plus input-reconstruction, label-inference, and deep-leakage experiments (PDF pp.~4--6, Algorithms~1--5 and Theorem~1; experimental Sec.~VI) & Identifiers, participation, model structure, timing, and any leakage outside the composed DP mechanism remain visible \\
\cite{quan2023hiersfl} & Intermediate results; perturbed local weights/updates; client/MES hierarchy metadata & Protect selected client releases through LDP & L=\AuditUnsupported{}; CG=\AuditUnsupported{}; DP=\AuditSupported{}; IA=\AuditUnsupported{}; XR=\AuditUnsupported{}. LDP applies to the stated weight/update release, not to label locality or the accumulated MES view & Formal LDP mechanism/argument and simulation (PDF pp.~3--5, Fig.~1, Algorithm~1, Sec.~II-C); no named inference-attack or cross-round analysis & Intermediate results, hierarchy membership, identities, timing, and repeated-round state outside the LDP release remain visible \\
\cite{stephanie2023dtasfl} & Activations; activation gradients; split-model parameters; secret shares; digital-twin status and staleness metadata & Protect selected aggregation values and assess reconstruction exposure & L=\AuditUnsupported{}; CG=\AuditSupported{}; DP=\AuditUnsupported{}; IA=\AuditSupported{}; XR=\AuditUnsupported{}. Secret sharing covers selected aggregation values, but labels/label-derived gradients and the accumulated cloud split-training view are not protected end-to-end & Analytical secret-sharing argument, reconstruction/privacy evaluation, and simulation (PDF pp.~4--10, Secs.~IV--V and Algorithms) & Cloud-visible activations, gradients, split state, hospital identity, timing, staleness, and repeated digital-twin metadata remain \\
\cite{meng2025pptadi} & Protected activations and labels; shares; model partitions; updates; TTP messages & Hide labels and selected intermediate values from the main server & L=\AuditSupported{}; CG=\AuditUnsupported{}; DP=\AuditUnsupported{}; IA=\AuditSupported{}; XR=\AuditUnsupported{}. The architectural label split/protection is supported, but the paper does not prove a general cryptographic confidentiality property for the delegate's accumulated view & Semi-honest threat analysis, prototype, and iDLG/model-inversion experiments (PDF pp.~5--6 and 10--12, Algorithms~1--3 and Sec.~4.3) & The TTP receives protected-operation inputs; model partitions, dimensions, identities, timing, and cross-round transcripts remain outside the demonstrated protection \\
\cite{stephanie2025weight} & Intermediate outputs; gradients; additive shares; initial-layer parameters; asynchronous status & Protect selected weights/updates in asynchronous SplitFed & L=\AuditUnsupported{}; CG=\AuditSupported{}; DP=\AuditUnsupported{}; IA=\AuditSupported{}; XR=\AuditUnsupported{}. Partial SMPC protects specified shared values, but it does not keep the full label-derived/cloud training view confidential & Analytical SMPC argument, reconstruction-oriented evaluation, and performance experiments (PDF pp.~7--20, Algorithms~1--3 and Sec.~4) & Activations, gradients, split state, participant identity, timing, and asynchronous history remain visible/linkable \\
\cite{zhao2023usfl} & Client-to-server smashed data; server-to-client transformed activation; activation gradients; split parameters & Keep labels, final classifier, and loss on the client & L=\AuditSupported{}; CG=\AuditUnsupported{}; DP=\AuditUnsupported{}; IA=\AuditUnsupported{}; XR=\AuditUnsupported{}. U-shaped placement gives architectural label locality but no general confidentiality guarantee for activations or accumulated server state & Architectural argument and prototype performance evaluation (PDF pp.~2--5, threat model and Algorithm~1); no named attack evaluation or cross-round analysis & The server observes per-client activations, gradients, middle-model state, order, timing, and repeated samples \\
\cite{chen2025llmiot} & Smashed data; activation gradients; LoRA modules/updates; chosen cut and resource metadata & Reduce representation leakage while meeting IoT resource constraints & L=\AuditNR{}; CG=\AuditUnsupported{}; DP=\AuditUnsupported{}; IA=\AuditLimited{}; XR=\AuditUnsupported{}. The Fisher-information proxy bounds/compares input reconstruction risk analytically but does not establish label locality, cryptographic confidentiality, DP, or accumulated-view privacy & Analytical Fisher-information leakage metric and simulation (PDF pp.~3--9, privacy metric, Algorithms~1--3, simulation Sec.~V) & Edge-visible activations, gradients, LoRA state, architecture, device identity, resource state, timing, and cross-round linkage remain \\
\cite{huang2025dpmora} & Activations; activation gradients; server-side submodel state; cut/resource-allocation metadata & Choose a cut subject to empirical reconstruction-risk constraints & L=\AuditUnsupported{}; CG=\AuditUnsupported{}; DP=\AuditUnsupported{}; IA=\AuditSupported{}; XR=\AuditUnsupported{}. Despite its name, DP-MORA supplies no differential-privacy mechanism; it measures and optimizes reconstruction risk & Real-data reconstruction experiments and real-device/performance measurements (PDF pp.~3--5 and 10--13, privacy model and experimental Sec.~VII) & The edge observes gradients, partition/model state, device identity, timing, repeated activations, and resource state; no formal leakage bound is given \\
\cite{chen2024shieldtse} & RSU temporal activations/embeddings; cloud outputs; gradients; updates & Keep traffic labels local and mitigate cloud-side input inversion & L=\AuditSupported{}; CG=\AuditUnsupported{}; DP=\AuditUnsupported{}; IA=\AuditSupported{}; XR=\AuditUnsupported{}. Labels remain at the RSU, while learned representations empirically resist two MI attacks; no formal confidentiality/DP or cross-round guarantee follows & Explicit honest-but-curious threat model and training-/optimization-based MI experiments (PDF pp.~5 and 8--11, Sec.~III-C and experimental Sec.~V) & The cloud accumulates embeddings, gradients, model state, RSU identity, timing, and repeated-round artifacts; empirical resistance is not cryptographic secrecy \\
\cite{hou2025verifyvfl} & Masked/encrypted local outputs and gradients; homomorphic signatures; proofs; batch metadata & Hide labels/sample-level gradient information from the computing party while making its result verifiable & L=\AuditSupported{}; CG=\AuditSupported{}; DP=\AuditUnsupported{}; IA=\AuditSupported{}; XR=\AuditUnsupported{}. The proved masking/signature view protects the specified label-related values under non-collusion, but does not analyze repeated-session linkability & Formal correctness/security proof and label-inference experiment with prototype measurements (PDF pp.~4--11, Algorithms, Sec.~VI Theorems~1--3, Sec.~VII) & Dimensions, batch linkage, proof transcripts, timing, authorized aggregates, and leakage beyond the theorem's view remain \\
\cite{latif2025hmac} & Activations; gradients; updates; HMAC tags; shared-key identifiers & No demonstrated executor-view confidentiality objective; authenticate split messages & L=\AuditNR{}; CG=\AuditUnsupported{}; DP=\AuditUnsupported{}; IA=\AuditUnsupported{}; XR=\AuditUnsupported{}. HMAC authenticates messages but neither hides labels/artifacts nor supplies privacy evidence & Analytical authenticity argument and tampering/poisoning prototype experiments (PDF pp.~3--11, Algorithm~1); privacy evaluation not reported & The delegate sees authenticated activations, gradients, updates, identifiers, sizes, ordering, timing, and repeated state \\
\cite{okegbile2025novel} & Semantic/smashed data; activation gradients; model updates; token comparisons; blockchain transactions and validation logs & Reduce semantic leakage and validate knowledge sharing & L=\AuditNR{}; CG=\AuditUnsupported{}; DP=\AuditUnsupported{}; IA=\AuditUnsupported{}; XR=\AuditUnsupported{}. Token-distance and blockchain mechanisms do not establish label locality, cryptographic confidentiality of the delegate's view, DP, or a named inference-attack guarantee & Analytical/qualitative privacy argument and performance evaluation (PDF pp.~4--12, Algorithm~1 and validation design); privacy-attack evidence not reported & Cloud-visible representations and gradients plus identities, token/log metadata, timing, and cross-round ledger linkage remain \\
\cite{cox2022aergia} & Offloaded model layers/parameters; signed schedules; device profiles; class-distribution similarity metadata & Keep labels at the weak client and protect selected scheduling metadata & L=\AuditSupported{}; CG=\AuditUnsupported{}; DP=\AuditUnsupported{}; IA=\AuditUnsupported{}; XR=\AuditUnsupported{}. Labels/loss remain with the weak client; SGX covers selected profiler/similarity processing, not the stronger peer's accumulated training view & Architectural/hardware argument and cluster prototype (PDF pp.~4--13, Secs.~3--5 and Algorithms~1--2); privacy-attack and cross-round evidence not reported & Strong peers see offloaded model/training state; schedules, identities, model portions, timing, and repeated assignments remain linkable \\
\cite{xia2022hsfl} & Client activations/intermediate results; labels; gradients; client/edge model parameters & Protect raw records through splitting, without a delegate-view confidentiality mechanism & L=\AuditUnsupported{}; CG=\AuditUnsupported{}; DP=\AuditUnsupported{}; IA=\AuditUnsupported{}; XR=\AuditUnsupported{}. The edge receives labels/intermediates and no privacy mechanism or attack analysis protects that accumulated view & Framework/algorithm plus performance experiment (PDF pp.~1--6, Sec.~II and Algorithm~1); privacy evidence not reported & Edge servers observe labels, activations, gradients, model state, client assignment, timing, and repeated rounds \\
\cite{wu2025hfsl} & Client activations/intermediate results; labels; auxiliary outputs; gradients; parameters & Protect raw records through heterogeneous splitting, without delegate-view secrecy & L=\AuditUnsupported{}; CG=\AuditUnsupported{}; DP=\AuditUnsupported{}; IA=\AuditUnsupported{}; XR=\AuditUnsupported{}. The split server receives label-related artifacts and no privacy mechanism/attack analysis is supplied & Framework/algorithms and server-class performance experiment (PDF pp.~5--22, Secs.~3--5 and Algorithms~1--3); privacy evidence not reported & The server accumulates labels or label-derived outputs, activations, gradients, model state, identities, timing, and repeated-round metadata \\
\cite{fei2025privacy} & Offloaded driver-data partition and task metadata; locally and remotely trained model parameters sent to cloud aggregation & Limit which driver data are offloaded & L=\AuditUnsupported{}; CG=\AuditUnsupported{}; DP=\AuditUnsupported{}; IA=\AuditUnsupported{}; XR=\AuditUnsupported{}. The RSU trains on the selected plaintext data partition, and no privacy mechanism protects that executor view & Analytical optimization and simulation (PDF pp.~4--13, Secs.~III--VII); no cryptographic/DP proof or privacy-attack evaluation & RSU observes the offloaded records/labels, task identity, order, timing, and resource state; cloud observes both update streams \\
\cite{he2023computation} & Offloaded training-data fraction; local/MEC model parameters; device/DT channel, energy, and allocation state & Optimize how much training data and computation move to MEC & L=\AuditUnsupported{}; CG=\AuditUnsupported{}; DP=\AuditUnsupported{}; IA=\AuditUnsupported{}; XR=\AuditUnsupported{}. The MEC server receives and trains on plaintext offloaded data, with no delegate-view privacy mechanism & System/optimization analysis and simulation on MNIST (PDF pp.~4--12, Secs.~II--IV and DDPG algorithm); no privacy proof or attack evaluation & MEC observes offloaded records/labels and model/resource state; identities, timing, repeated allocation history, and updates remain visible \\
\end{longtable}

\begin{longtable}{
  >{\raggedright\arraybackslash}p{0.055\linewidth}
  >{\raggedright\arraybackslash}p{0.195\linewidth}
  >{\raggedright\arraybackslash}p{0.30\linewidth}
  >{\raggedright\arraybackslash}p{0.18\linewidth}
  >{\raggedright\arraybackslash}p{0.21\linewidth}
}
\caption{Delegated-attribution support and evidence. A mechanism receives credit only for the links it provides.}
\label{tab:audit_integrity}\\
\toprule
\textbf{Ref.} & \textbf{Attribution objective} & \textbf{Design support} & \textbf{Evidence status} & \textbf{Verifier / assumption} \\
\midrule
\endfirsthead
\toprule
\textbf{Ref.} & \textbf{Attribution objective} & \textbf{Design support} & \textbf{Evidence status} & \textbf{Verifier / assumption} \\
\midrule
\endhead
\bottomrule
\endfoot
\cite{dong2020offloading} & Correct delegated training inside the TEE & \AuditLimited{}: hardware isolation is intended to support correctness inside the enclave; completeness, authorized-input binding, freshness, and replay resistance are not separately provided & Hardware assumption; simulation only; attestation evidence \AuditNR{} & TEE/attestation infrastructure; verifier details not reported \\
\cite{jijin2022smart} & Authenticate/provenance-track delegated tasks and outputs & \AuditLimited{}: transactions and logs support provenance and process consistency; computation correctness, completeness, input binding, freshness, and replay resistance are \AuditUnsupported{} & Analytical argument; prototype measurement & Smart contracts/blockchain under consensus and logging assumptions \\
\cite{zhang2025zkfhed} & Verify request, encrypted training, and decryption relations & \AuditSupported{} for the relations represented by the circuits; \AuditLimited{} for full-work completeness, dataset authorization, and cross-round freshness not encoded in the statement & Formal security proof; prototype measurement & Cryptographic verifier under FHE/ZK setup and key assumptions \\
\cite{han2025outsourced} & Detect inconsistent or evasive outsourced training behavior & \AuditLimited{}: sensitive-sample checks support behavioral output consistency; full correctness, completeness, input binding, freshness, and replay resistance are \AuditUnsupported{} & Empirical evaluation; prototype measurement & Data owner under a statistical/behavioral audit assumption \\
\cite{lin2022td3} & Verify selected edge/fog execution & \AuditUnsupported{}: no computation-verification mechanism is provided & Integrity evidence \AuditNR{}; simulation only for performance & No verifier \\
\cite{marx2023ww} & Ensure correct private cluster training and secure global aggregation over the current round's secret-shared model and data & \AuditSupported{} for the Train, Reshare, and Agg relations realized by Algorithms~1 and~3 under Theorem~1; \AuditLimited{} for owner-input validity, permitted sample policy, and explicit replay or contribution-admission semantics outside the ideal functionality & Formal simulation-based security proof; CrypTen/AWS prototype measurement & MPC parties and configured corruption threshold; no separate external verifier \\
\cite{chadha2020towards} & Verify FaaS-executed local training and admit the resulting client update & \AuditLimited{} for runtime completion/error handling, retry, and current model/optimizer-state forwarding; prescribed-work correctness is \AuditUnsupported{} because client-code consistency is explicitly assumed, with no verification of samples, epochs, optimizer use, freshness, or replay & Prototype measurement; prescribed-work and adversarial-executor evaluation \AuditNR{} & Weights-Updater and FaaS runtime under a correct-code/execution assumption; no computation verifier \\
\cite{li2021quantum} & Establish correct quantum-server evaluation of the authorized per-client gradient before shared-model updating & \AuditLimited{}: the paper invokes UBQC security/integrity and the client interactively derives the result, but it specifies no independent prescribed-work check, current-round/replay binding, or failed-verification consequence for aggregation & Analytical claim inherited from UBQC; numerical performance and privacy-attack simulation only; adversarial-computation verification evidence \AuditNR{} & Client supplies adaptive measurement instructions and derives the result; no explicit verifier or admission rule \\
\cite{grafberger2021fedless} & Authenticate authorized client-function invocation and admit completed local updates & \AuditLimited{}: JWT/scoped credentials authenticate requests, clients retrieve the latest model, and timeout/failure status gates returned updates; computation correctness, required work, contribution-to-round binding, and replay resistance are not established & Security architecture and cross-cloud prototype measurement; computation-verification evidence \AuditNR{} & Trusted Cognito signing key, controller, and parameter server; no executor attestation or computation verifier \\
\cite{elzohairy2022fedlesscan} & Bind delayed client updates to their training round and prevent obsolete contributions from affecting aggregation & \AuditSupported{} for the explicit update-age test and discard rule; prescribed-work correctness, completeness, authorized sample-set/epoch binding, and malicious-executor resistance are \AuditUnsupported{} or \AuditNR{} & Analytical algorithmic argument; GCF prototype measurement with simulated stragglers & Controller/aggregator and client-reported history under a non-malicious-client assumption; no computation verifier \\
\cite{xu2025sgboost} & Establish current-task execution and admission for protected tree-training subroutines & Fr=\AuditUnsupported{}; W=\AuditUnsupported{}; AD=\AuditUnsupported{}. Encryption/share consistency under honest-but-curious execution is authentication/confidentiality context, not proof of the authorized samples, parameters, tree step, completeness, or malicious execution; no verification-gated admission exists & Security analysis and prototype performance only (PDF pp.~4 and 9--13); no empirical integrity attack or prescribed-work verifier & Protocol parties under non-collusion; no malicious executor verifier, nonce/round check, or contribution-admission rule \\
\cite{thapa2022splitfed} & Verify main-server forward/backward execution & Fr=\AuditUnsupported{}; W=\AuditUnsupported{}; AD=\AuditUnsupported{}. Algorithms exchange the expected tensors but do not bind them cryptographically to the current authorized state or verify that the main server performed the prescribed layers/steps; no admission consequence is specified & Protocol algorithms and performance experiments (PDF pp.~3--7, Algorithms~1--2); adversarial-execution evidence not reported & No computation verifier or admission controller \\
\cite{wu2023federated} & Verify complementary submodel computation & Fr=\AuditUnsupported{}; W=\AuditUnsupported{}; AD=\AuditUnsupported{}. DP and contract/incentive mechanisms do not establish current-state binding, correct/completed complementary learning, or verification-triggered aggregation control & Algorithms/theory and privacy/incentive simulation (PDF pp.~4--10); no integrity attack or computation proof & No computation verifier \\
\cite{quan2023hiersfl} & Verify MES/cloud split training & Fr=\AuditUnsupported{}; W=\AuditUnsupported{}; AD=\AuditUnsupported{}. The hierarchical protocol supplies no authenticated current-state binding, prescribed-work proof, or verification-linked admission rule & Framework algorithm and simulation (PDF pp.~3--5, Algorithm~1); integrity evidence not reported & No verifier \\
\cite{stephanie2023dtasfl} & Bind asynchronous updates to step/staleness state and condition their aggregation & Fr=\AuditSupported{}; W=\AuditUnsupported{}; AD=\AuditSupported{}. Step-aware/staleness metadata binds returned updates to the asynchronous model state and explicitly changes their aggregation weight/admission, but secret sharing does not verify cloud-side forward/backward work or completeness & Algorithmic/analytical argument and simulation (PDF pp.~4--10, Sec.~IV and update-weight equations) & Cloud/asynchronous aggregator is the freshness/admission decision point; no delegated-training verifier \\
\cite{meng2025pptadi} & Verify protected operations and main-server split training & Fr=\AuditNR{}; W=\AuditNR{}; AD=\AuditNR{}. Semi-honest protocol execution and successful completion do not by themselves establish current-task binding, policy-relative prescribed work, or an aggregation consequence, and the paper gives insufficient evidence to code those properties & Algorithms, threat model, attack experiments, and prototype (PDF pp.~5--12); computation-verification/admission semantics not reported & TTP/main server under semi-honest assumptions; no explicit prescribed-work verifier \\
\cite{stephanie2025weight} & Bind asynchronous updates to model/version state and condition their aggregation & Fr=\AuditSupported{}; W=\AuditUnsupported{}; AD=\AuditSupported{}. Version/step-aware asynchronous weighting binds updates to current state and affects aggregation, but partial SMPC only protects shared values and does not verify the cloud training segment & Algorithmic/analytical argument and experiments (PDF pp.~7--20, Algorithms~1--3 and weighting equations) & Cloud/asynchronous aggregator applies the consequence; no delegated-computation verifier \\
\cite{zhao2023usfl} & Verify split-server middle-segment execution & Fr=\AuditNR{}; W=\AuditNR{}; AD=\AuditNR{}. The workflow specifies tensor exchange but does not report a freshness mechanism, prescribed-work verification, or a verification-linked admission rule & Architecture/Algorithm~1 and performance experiment (PDF pp.~2--5); integrity evidence not reported & No verifier identified \\
\cite{chen2025llmiot} & Verify edge LoRA/upper-model computation & Fr=\AuditNR{}; W=\AuditNR{}; AD=\AuditNR{}. Resource/cut optimization does not report current-state binding, execution verification, or a consequence for admitting the owner-attributed result & Analytical optimization and simulation (PDF pp.~3--9, Algorithms~1--3); integrity evidence not reported & No verifier identified \\
\cite{huang2025dpmora} & Verify edge-side submodel computation & Fr=\AuditNR{}; W=\AuditNR{}; AD=\AuditNR{}. Reconstruction-risk estimation and cut/resource selection do not establish that the edge performed the prescribed current training work or define verification-gated admission & Analytical optimization, reconstruction experiments, and device measurements (PDF pp.~3--13); integrity evidence not reported & Leakage estimator is not a computation verifier \\
\cite{chen2024shieldtse} & Verify cloud-side representation/training computation & Fr=\AuditNR{}; W=\AuditNR{}; AD=\AuditNR{}. The honest-but-curious model assumes correct training and reports neither freshness binding, prescribed-work proof, nor an admission consequence & Privacy-attack and performance experiments (PDF pp.~5--11); integrity evidence not reported & No verifier identified \\
\cite{hou2025verifyvfl} & Verify the outsourced gradient relations and reject invalid results & Fr=\AuditUnsupported{}; W=\AuditSupported{}; AD=\AuditSupported{}. Homomorphic signatures/proofs establish the specified computation relations and failed verification prevents use of the result; the proof does not bind the transcript to a current round/session/nonce or establish replay freshness & Formal correctness/security proof and prototype integrity measurements (PDF pp.~4--11, algorithms, Sec.~VI Theorems~1--3, Sec.~VII) & Data owners verify under TA-key and non-collusion assumptions; work assurance is limited to the proved batch-level relations \\
\cite{latif2025hmac} & Authenticate tensors/updates and discard tampered contributions & Fr=\AuditUnsupported{}; W=\AuditUnsupported{}; AD=\AuditSupported{}. HMAC verifies origin/message integrity and failures cause activations or poisoned updates to be discarded, but it does not prove the learning transformation, samples, epochs, optimizer, completeness, or freshness & Analytical authenticity argument and tampering/poisoning experiment (PDF pp.~3--11, Algorithm~1 and discard rules) & Endpoints share the HMAC key; authentication is not computation verification \\
\cite{okegbile2025novel} & Validate model/knowledge-sharing artifacts before DT evolution & Fr=\AuditNR{}; W=\AuditNR{}; AD=\AuditSupported{}. Token/blockchain validation has an explicit accept/reject consequence for model evolution, but current-state binding and policy-relative correctness/completeness of cloud-side learning are insufficiently specified & Analytical validation design and performance evaluation (PDF pp.~5--12, Algorithm~1 and blockchain-validation equations) & Blockchain validators/authority under incompletely reported trust and key assumptions \\
\cite{cox2022aergia} & Bind peer-assisted work to the scheduled FL round & Fr=\AuditSupported{}; W=\AuditUnsupported{}; AD=\AuditUnsupported{}. Signed schedules and round/deadline handling identify the current assignment, but they do not verify that the stronger peer executed all prescribed layers/steps; scheduling or late-result handling is not a verification-triggered admission guarantee & Architectural/hardware argument and cluster prototype (PDF pp.~5--13, Sec.~4 and Algorithms~1--2) & Federator/SGX handle scheduling metadata; no peer-computation verifier \\
\cite{xia2022hsfl} & Verify edge-side split training & Fr=\AuditNR{}; W=\AuditNR{}; AD=\AuditNR{}. The paper reports the collaborative workflow but insufficient evidence of freshness, prescribed-work verification, or verification-linked admission & Framework/Algorithm~1 and performance evaluation (PDF pp.~2--6); integrity evidence not reported & No verifier identified \\
\cite{wu2025hfsl} & Verify split-server training & Fr=\AuditNR{}; W=\AuditNR{}; AD=\AuditNR{}. The distributed protocol lacks reported current-state binding, prescribed-work verification, and a verification-linked admission rule & Framework/Algorithms~1--3 and server-class experiment (PDF pp.~5--22); integrity evidence not reported & No verifier identified \\
\cite{fei2025privacy} & Verify RSU training of the authorized offloaded data partition & Fr=\AuditUnsupported{}; W=\AuditUnsupported{}; AD=\AuditUnsupported{}. The SAC/offloading policy selects a data amount and resource action, but supplies no authentication/freshness, computation proof, or verification-based aggregation consequence & Optimization analysis and simulation (PDF pp.~4--13, Secs.~III--VII); integrity evidence not reported & No verifier; cooperative RSU/cloud execution is assumed \\
\cite{he2023computation} & Verify MEC training of the authorized offloaded data fraction & Fr=\AuditUnsupported{}; W=\AuditUnsupported{}; AD=\AuditUnsupported{}. DDPG chooses the offloading/resource policy, but no mechanism proves the MEC used the authorized data/model/steps or binds and gates the returned result & Optimization analysis and simulation (PDF pp.~4--12, Secs.~II--IV and DDPG algorithm); integrity evidence not reported & No verifier; cooperative MEC/aggregation execution is assumed \\
\end{longtable}

\begin{longtable}{
  >{\raggedright\arraybackslash}p{0.055\linewidth}
  >{\raggedright\arraybackslash}p{0.20\linewidth}
  >{\raggedright\arraybackslash}p{0.275\linewidth}
  >{\raggedright\arraybackslash}p{0.19\linewidth}
  >{\raggedright\arraybackslash}p{0.22\linewidth}
}
\caption{Residual owner burden and constrained-owner feasibility. Design support describes cost relocation; evidence status describes the measurements actually reported.}
\label{tab:audit_client_burden}\\
\toprule
\textbf{Ref.} & \textbf{Owner-side objective} & \textbf{Design support} & \textbf{Evidence status} & \textbf{Residual costs / inapplicable components} \\
\midrule
\endfirsthead
\toprule
\textbf{Ref.} & \textbf{Owner-side objective} & \textbf{Design support} & \textbf{Evidence status} & \textbf{Residual costs / inapplicable components} \\
\midrule
\endhead
\bottomrule
\endfoot
\cite{dong2020offloading} & Reduce client training computation through full TEE offloading & \AuditLimited{}: local training is removed, but encryption, packaging, upload, setup, and attestation-related work remain & Simulation only for deadline/frequency; client RAM, energy, setup, and verification measurements \AuditNR{} & Local training: removed; proof generation: \AuditNA{}; encryption/communication/setup remain \\
\cite{jijin2022smart} & Reduce client execution through service-node offloading & \AuditLimited{}: training is delegated, but raw-data upload, transaction handling, and log state remain & Prototype measurement on physical v-FAP infrastructure; client energy, RAM, and security overhead \AuditNR{} & Communication and persistent ledger/protocol state remain; proof generation: \AuditNA{} \\
\cite{zhang2025zkfhed} & Remove plaintext local training while preserving encrypted execution and verifiability & \AuditLimited{}: training is outsourced, but encryption, request construction, communication, proof verification, and decryption participation remain & Prototype measurement; embedded real-device RAM and energy evaluation \AuditNR{} & Ciphertext/proof expansion, setup, key state, and verification remain \\
\cite{han2025outsourced} & Reduce local training through private-cloud outsourcing & \AuditLimited{}: training is outsourced, but batch upload and sensitive-sample audit work remain & Prototype measurement of audit overhead; constrained-device RAM, energy, and communication evidence \AuditNR{} & Preprocessing/audit state and communication remain; proof generation: \AuditNA{} \\
\cite{lin2022td3} & Reduce device delay/energy through edge/fog selection & \AuditLimited{}: delegated execution reduces local work but retains task upload, synchronization, and selection state & Simulation only; real-device RAM and security measurements \AuditNR{} & Communication and policy state remain; proof generation: \AuditNA{} \\
\cite{marx2023ww} & Remove owner-side model training by outsourcing PPML training to regional MPC clusters & \AuditLimited{}: clients only secret-share selected data and may then leave, but input preparation, share generation/distribution, trust-zone selection, and later data contribution remain & CrypTen/AWS server-side prototype and WAN measurement; constrained-client RAM, energy, and share-generation measurements \AuditNR{} & Client data preparation, secret-sharing setup, communication, trust-configuration choice, and possible cross-round data submission remain; local model storage and training are removed \\
\cite{chadha2020towards} & Remove owner-side local optimization through commercial FaaS client execution & \AuditLimited{}: the GCF branch removes local training, but owners retain data staging, task/function configuration, communication, invocation credentials, and client/optimizer state; self-hosted branches retain infrastructure responsibility & Prototype measurement across GCF, OpenWhisk, and OpenFaaS; constrained-owner RAM, energy, setup, and security-overhead measurements \AuditNR{} & Cloud-bucket/data setup, model configuration, authentication, communication, and persistent optimizer state remain; proof generation: \AuditNA{} \\
\cite{li2021quantum} & Use a remote quantum server for expensive loss/gradient evaluation without requiring a client quantum computer & \AuditLimited{}: the server performs the delegated quantum gradient computation, but each client prepares and sends single qubits, conducts adaptive two-way classical interaction, derives the result, and clips/adds noise; Algorithm~1 also includes Adam/model updating without identifying a separate aggregator & Numerical simulation only on MNIST and WDBC; no real quantum-device, client RAM, energy, communication, or recovery measurement & Single-qubit or encoded-state preparation, interactive measurement-control state, communication, result derivation, gradient clipping/noise, optimizer state, and model-update handling remain; amplitude-state preparation may itself be inefficient \\
\cite{grafberger2021fedless} & Remove owner-side local optimization while scaling authenticated FaaS clients & \AuditLimited{}: commercial-FaaS execution removes local training, but the data holder deploys/registers the function, supplies data locations and optional platform secrets, communicates updates, and maintains external LDP-budget state when used & Cross-cloud prototype measurement with up to 200 client functions and an IaaS cost/performance comparison; constrained-device RAM, energy, setup, and verification measurements \AuditNR{} & Function deployment, registration, storage, authentication secrets, communication, privacy-budget state, and recovery from failed invocations remain; proof generation: \AuditNA{} \\
\cite{elzohairy2022fedlesscan} & Reduce owner computation and straggler cost through GCF-hosted client training & \AuditLimited{}: GCF functions perform local optimization, but data/function setup, FedLess registration and authentication, communication, and persistent client-history state remain & GCF prototype measurement on four datasets with measured time/cost and simulated straggler ratios; constrained-owner RAM, energy, setup, and security-overhead measurements \AuditNR{} & Data storage, client-function configuration, authentication, communication, timing/missed-round history, and failed-update recovery remain; proof generation: \AuditNA{} \\
\cite{xu2025sgboost} & Distribute protected vertical tree-training work & LW=\AuditSupported{}; BX=\AuditSupported{}; AW=\AuditSupported{}; IoT=\AuditUnsupported{}; M+C=\AuditUnsupported{}. Passive owners retain bucketization/feature-side operations, cryptographic preparation, and interactive combination of returned shares/ciphertexts & Desktop/laptop prototype with computation/communication measurements (PDF pp.~10--13, Sec.~VII); no constrained-IoT execution and no joint peak owner-memory plus communication report & Recurring cryptographic preprocessing, randomness, interaction, model state, and communication remain; setup/key state is persistent \\
\cite{thapa2022splitfed} & Reduce client model execution through split learning & LW=\AuditSupported{}; BX=\AuditSupported{}; AW=\AuditSupported{}; IoT=\AuditUnsupported{}; M+C=\AuditUnsupported{}. Clients retain designated forward/backward layers, loss work in label-local variants, DP perturbation where used, and combine returned activation gradients with local backpropagation & Prototype timing/communication and model-size analysis (PDF pp.~3--7); no complete constrained-device workflow or joint peak memory/communication evidence & Per-batch smashed-data/gradient exchange, local model/optimizer state, update synchronization, and optional DP cost remain \\
\cite{wu2023federated} & Distribute complementary data- and label-side submodel computation & LW=\AuditSupported{}; BX=\AuditSupported{}; AW=\AuditSupported{}; IoT=\AuditUnsupported{}; M+C=\AuditUnsupported{}. Each owner executes a necessary submodel, combines returned gradients, and performs clipping/perturbation plus identifier-linked exchange & Simulation/performance and privacy-attack measurements (PDF pp.~8--11, Sec.~VI); no constrained-vehicle hardware workflow and no joint peak owner-memory/communication report & Local forward/backward work, DP preprocessing, identifiers, synchronization, and repeated communication remain \\
\cite{quan2023hiersfl} & Reduce owner burden through MES/cloud hierarchy & LW=\AuditSupported{}; BX=\AuditSupported{}; AW=\AuditSupported{}; IoT=\AuditUnsupported{}; M+C=\AuditUnsupported{}. Clients retain local layers/updates, receive edge-side results for completion, and add LDP perturbation & Simulation on learning/communication behavior (PDF pp.~4--6, Sec.~III); no real mobile/IoT device and no joint peak memory/communication measurement & Local learning, perturbation, repeated client--MES exchange, optimizer/model state, and synchronization remain \\
\cite{stephanie2023dtasfl} & Reduce IoMT burden through digital-twin-assisted asynchronous splitting & LW=\AuditSupported{}; BX=\AuditSupported{}; AW=\AuditSupported{}; IoT=\AuditUnsupported{}; M+C=\AuditUnsupported{}. Hospital/edge components retain shallow-model work, combine cloud-returned gradients, create shares, and maintain step/digital-twin state & Simulation/experimental configuration on modeled hospitals (PDF pp.~9--10, Sec.~V); no complete constrained-IoMT device workflow and no joint peak memory/communication report & Per-batch activations/gradients, shares, local optimizer/model state, staleness metadata, and recovery/synchronization remain \\
\cite{meng2025pptadi} & Reduce AIoT training while protecting U-shaped intermediates & LW=\AuditSupported{}; BX=\AuditSupported{}; AW=\AuditSupported{}; IoT=\AuditUnsupported{}; M+C=\AuditUnsupported{}. Devices retain input/output model parts and loss/backpropagation, combine returned features/gradients, and perform protected-operation messaging & Prototype and latency/attack measurements (PDF pp.~10--12); no constrained-IoT end-device workflow and no joint peak memory/communication report & Two local model segments, crypto/sharing messages, synchronization, keys/shares, and recurring communication remain \\
\cite{stephanie2025weight} & Reduce owner burden through weighted asynchronous SplitFed & LW=\AuditSupported{}; BX=\AuditSupported{}; AW=\AuditSupported{}; IoT=\AuditUnsupported{}; M+C=\AuditUnsupported{}. Owner/edge components retain local layers, combine returned gradients, generate shares, and maintain asynchronous version state & Multimedia-health experiments and communication/latency measurements (PDF pp.~14--20, Sec.~4); no complete constrained-IoT device workflow and no joint peak memory/communication report & Local layers, SMPC/share state, version metadata, activation/gradient traffic, and reconnection/staleness handling remain \\
\cite{zhao2023usfl} & Reduce client computation while retaining labels and loss locally & LW=\AuditSupported{}; BX=\AuditSupported{}; AW=\AuditUnsupported{}; IoT=\AuditUnsupported{}; M+C=\AuditUnsupported{}. Clients retain first/last model segments, loss and backward completion, and combine the returned middle-segment activation; no added owner-side privacy/integrity mechanism is credited & Prototype/performance experiments (PDF pp.~3--5); no complete constrained-device workflow and no joint peak memory/communication report & Two model segments, optimizer state, two-boundary activation/gradient traffic, synchronization, and per-round state remain \\
\cite{chen2025llmiot} & Fit split LoRA/LLM fine-tuning to heterogeneous IoT resource classes & LW=\AuditSupported{}; BX=\AuditSupported{}; AW=\AuditUnsupported{}; IoT=\AuditUnsupported{}; M+C=\AuditUnsupported{}. Devices retain selected layers/LoRA computation and combine returned activation gradients; the resource optimizer is not credited as privacy/integrity work & Analytical cost model and simulation using named device profiles (PDF pp.~3--9, Secs.~III--V); no actual constrained-device run and no joint measured peak memory/communication & Local LoRA/model state, cut configuration, activation/gradient traffic, bandwidth/energy demand, and synchronization remain \\
\cite{huang2025dpmora} & Select a latency/leakage-aware split and resource allocation & LW=\AuditSupported{}; BX=\AuditSupported{}; AW=\AuditSupported{}; IoT=\AuditSupported{}; M+C=\AuditUnsupported{}. Devices retain the lower submodel, combine returned gradients, estimate reconstruction risk, and track cut/resource state & Real Raspberry Pi 3B, 3A+, and 4B measurements plus real-dataset reconstruction/performance experiments (PDF pp.~10--13, Sec.~VII); joint peak memory and communication are not both reported & Local forward/backward work, privacy-risk estimation, activation/gradient traffic, resource state, optimizer/model state, and recurring energy/latency costs remain \\
\cite{chen2024shieldtse} & Reduce RSU computation while mitigating representation inversion & LW=\AuditSupported{}; BX=\AuditSupported{}; AW=\AuditSupported{}; IoT=\AuditUnsupported{}; M+C=\AuditUnsupported{}. RSUs retain temporal encoder/head and loss/backpropagation, combine returned cloud gradients, and run the privacy-oriented representation objective & Server-class GPU prototype/privacy measurements (PDF pp.~8--11, Sec.~V); no RSU-class constrained device and no joint peak owner-memory/communication report & Encoder/head parameters, local optimizer, privacy-loss computation, embedding/gradient traffic, synchronization, and state remain \\
\cite{hou2025verifyvfl} & Outsource VFL gradient computation with owner verification & LW=\AuditSupported{}; BX=\AuditSupported{}; AW=\AuditSupported{}; IoT=\AuditUnsupported{}; M+C=\AuditUnsupported{}. Owners compute local outputs/masks/signatures, verify batch proofs, and combine verified returned results with local updates & Prototype proof-generation, verification, and communication measurements (PDF pp.~9--11, Sec.~VII); no constrained-IoT execution and no joint peak memory/communication report & Per-batch crypto, proof verification, key/signature state, local model/optimizer work, and expanded communication remain \\
\cite{latif2025hmac} & Add low-cost authenticity to split messages & LW=\AuditSupported{}; BX=\AuditSupported{}; AW=\AuditSupported{}; IoT=\AuditUnsupported{}; M+C=\AuditUnsupported{}. CIoT clients retain split layers, combine returned gradients, and generate/verify HMAC tags under shared-key state & Prototype attack-detection and timing measurements (PDF pp.~4--11); no complete constrained-IoT hardware workflow and no joint peak memory/communication report & Local forward/backward work, tags, key setup/storage, per-message verification, activation/gradient traffic, and retry/discard handling remain \\
\cite{okegbile2025novel} & Reduce physical-twin burden while adding semantic defense and validation & LW=\AuditSupported{}; BX=\AuditSupported{}; AW=\AuditSupported{}; IoT=\AuditUnsupported{}; M+C=\AuditUnsupported{}. The owner/edge side retains semantic extraction and split-model work, combines cloud-returned gradients, and produces token/validation and blockchain-related state & Analytical cost model and performance evaluation (PDF pp.~9--16, optimization and Sec.~VI); no constrained physical-twin device and no joint peak memory/communication report & Semantic preprocessing, local model/optimizer state, smashed-data/gradient traffic, token comparisons, logs, validation energy, and persistent ledger metadata remain \\
\cite{cox2022aergia} & Shift selected layers from weak to stronger peers & LW=\AuditSupported{}; BX=\AuditUnsupported{}; AW=\AuditSupported{}; IoT=\AuditUnsupported{}; M+C=\AuditUnsupported{}. Weak clients retain necessary local layer training and profiling/schedule handling, but delegated peer results flow to the federator rather than returning for owner-local combination & Cluster prototype and performance measurements (PDF pp.~9--13, Sec.~5); no target-IoT device and no joint peak owner-memory/communication report & Local unfrozen layers, profiling, signed schedules, model upload, assignment state, and synchronization remain; the federator, not the weak owner, recombines model portions \\
\cite{xia2022hsfl} & Reduce client execution through hierarchical split training & LW=\AuditSupported{}; BX=\AuditSupported{}; AW=\AuditUnsupported{}; IoT=\AuditUnsupported{}; M+C=\AuditUnsupported{}. Clients retain local forward/backward layers and combine returned edge gradients; no additional owner-side privacy/integrity mechanism is credited & Performance experiment/emulation on real datasets (PDF pp.~4--6); no constrained-IoT hardware workflow and no joint peak memory/communication report & Local layers/optimizer state, label/intermediate transfer, gradient receipt, model synchronization, and repeated traffic remain \\
\cite{wu2025hfsl} & Adapt split training to heterogeneous client models & LW=\AuditSupported{}; BX=\AuditSupported{}; AW=\AuditUnsupported{}; IoT=\AuditUnsupported{}; M+C=\AuditUnsupported{}. Clients retain selected model work and combine returned server artifacts; no additional owner-side privacy/integrity mechanism is credited & Server-class prototype/performance experiment (PDF pp.~20--22, Sec.~5); no constrained-client execution and no joint peak memory/communication report & Local model/optimizer state, label/activation/auxiliary-output and gradient traffic, adapter state, and repeated synchronization remain \\
\cite{fei2025privacy} & Split training between vehicle-local and RSU-executed data partitions & LW=\AuditSupported{}; BX=\AuditUnsupported{}; AW=\AuditUnsupported{}; IoT=\AuditUnsupported{}; M+C=\AuditUnsupported{}. Vehicles retain necessary local training, but the RSU-trained branch goes to cloud aggregation rather than returning for owner-local completion; no added privacy/integrity mechanism is credited & Optimization analysis and simulation (PDF pp.~4--13, Secs.~III--VII); no vehicle/IoT training experiment and no joint peak memory/communication report & Local training/model state, offloaded-data upload for the delegated branch, resource-policy state, update transmission, latency, energy, and global-model synchronization remain \\
\cite{he2023computation} & Split training between device-local and MEC-executed data fractions & LW=\AuditSupported{}; BX=\AuditUnsupported{}; AW=\AuditUnsupported{}; IoT=\AuditUnsupported{}; M+C=\AuditUnsupported{}. The mobile device trains its retained data fraction, while the MEC-trained branch is sent to aggregation rather than returned for owner-local combination; no added privacy/integrity mechanism is credited & DDPG optimization and MNIST simulation (PDF pp.~9--12, Sec.~IV); no constrained mobile-device experiment and no joint peak memory/communication report & Local training/model state, plaintext data upload for the delegated fraction, DT/resource state, update transmission, latency/energy, and global synchronization remain \\
\end{longtable}

\end{document}